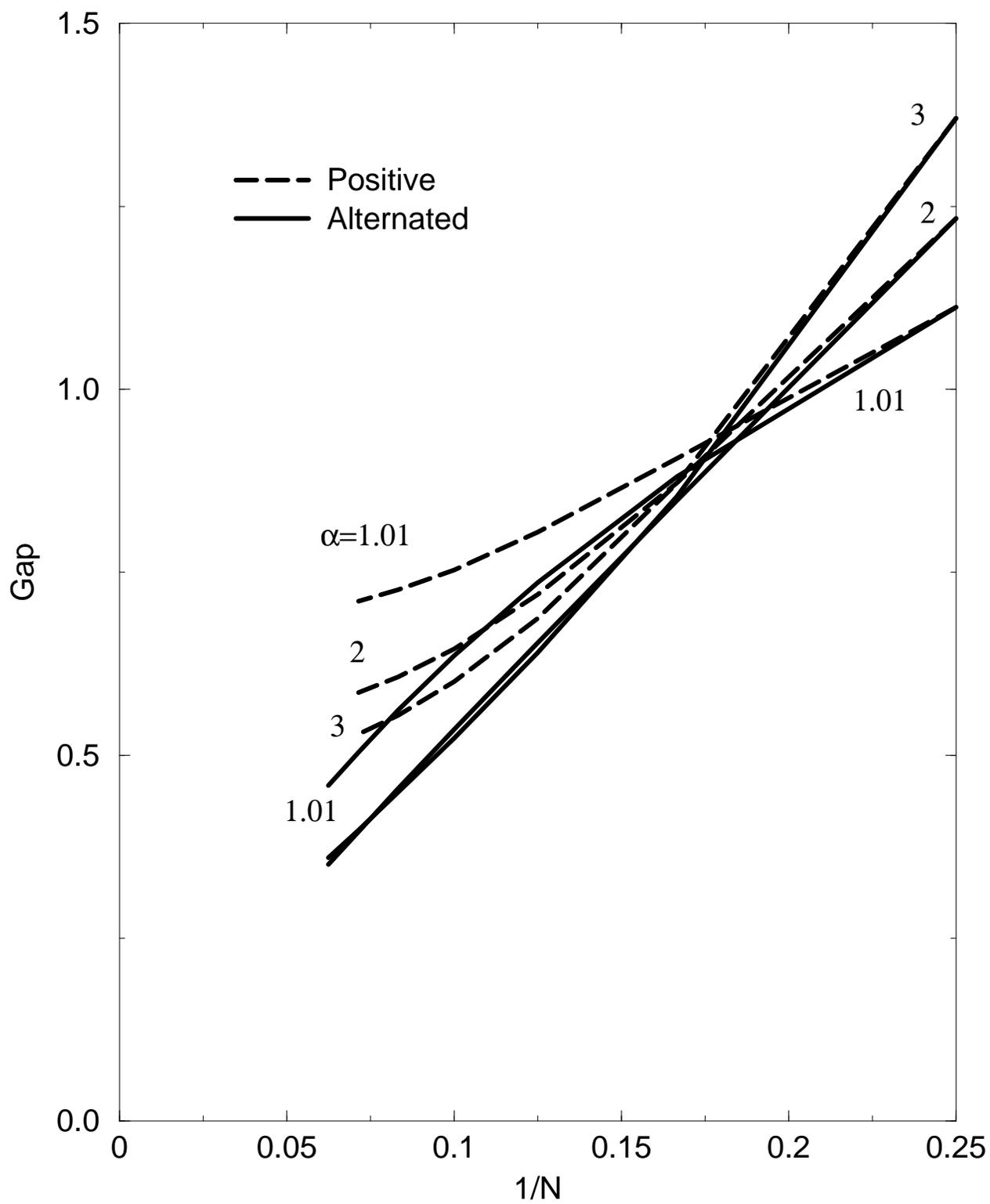

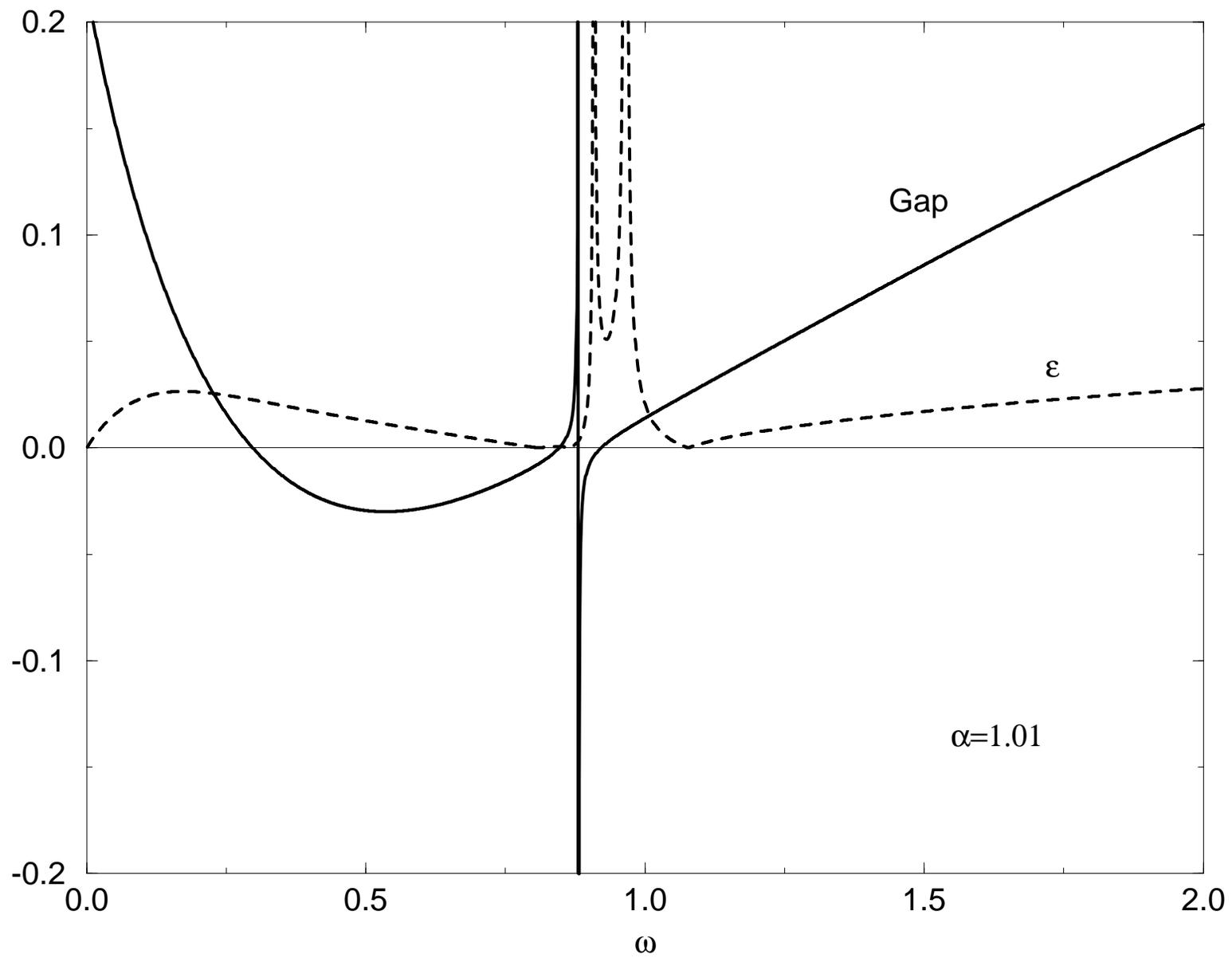

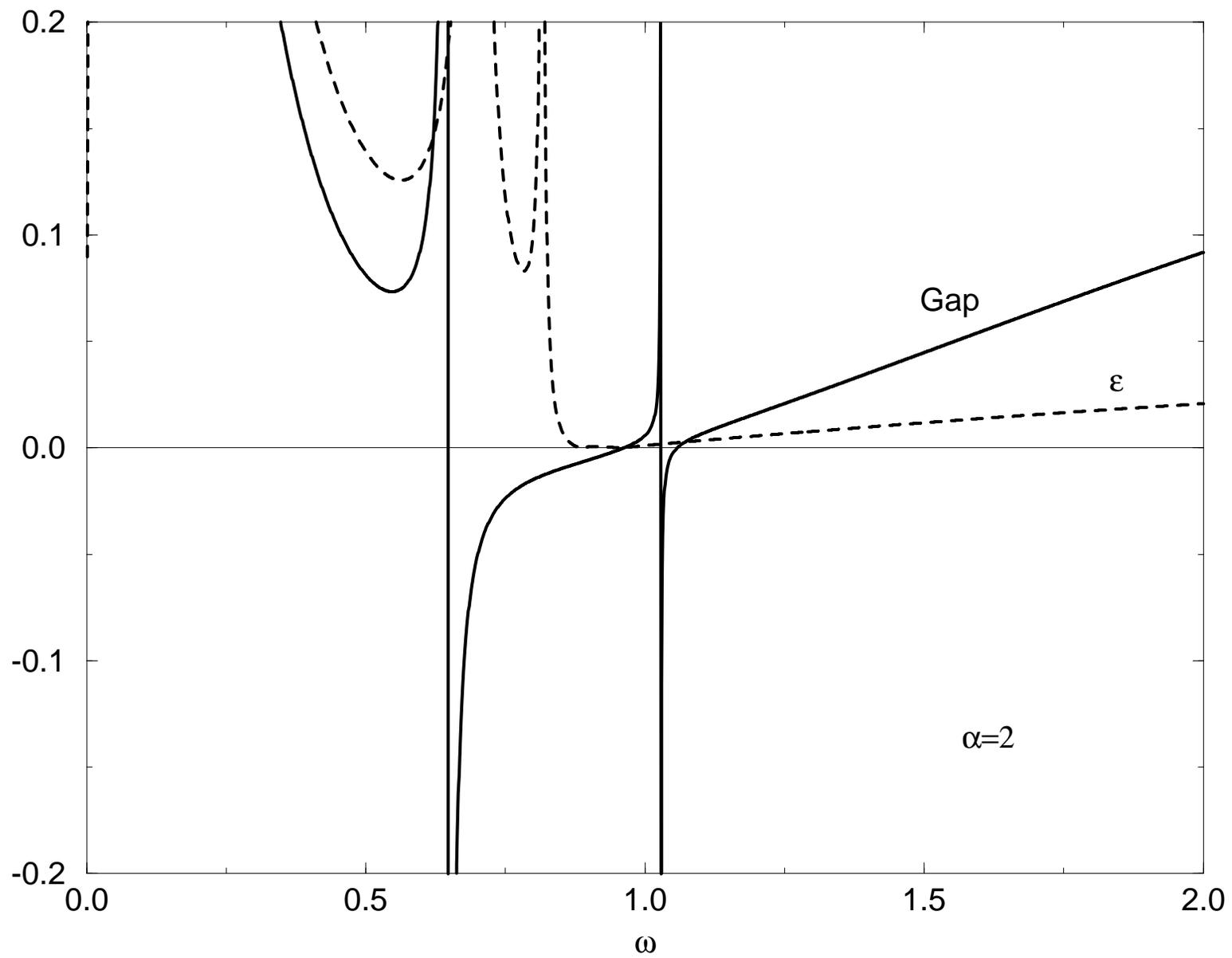

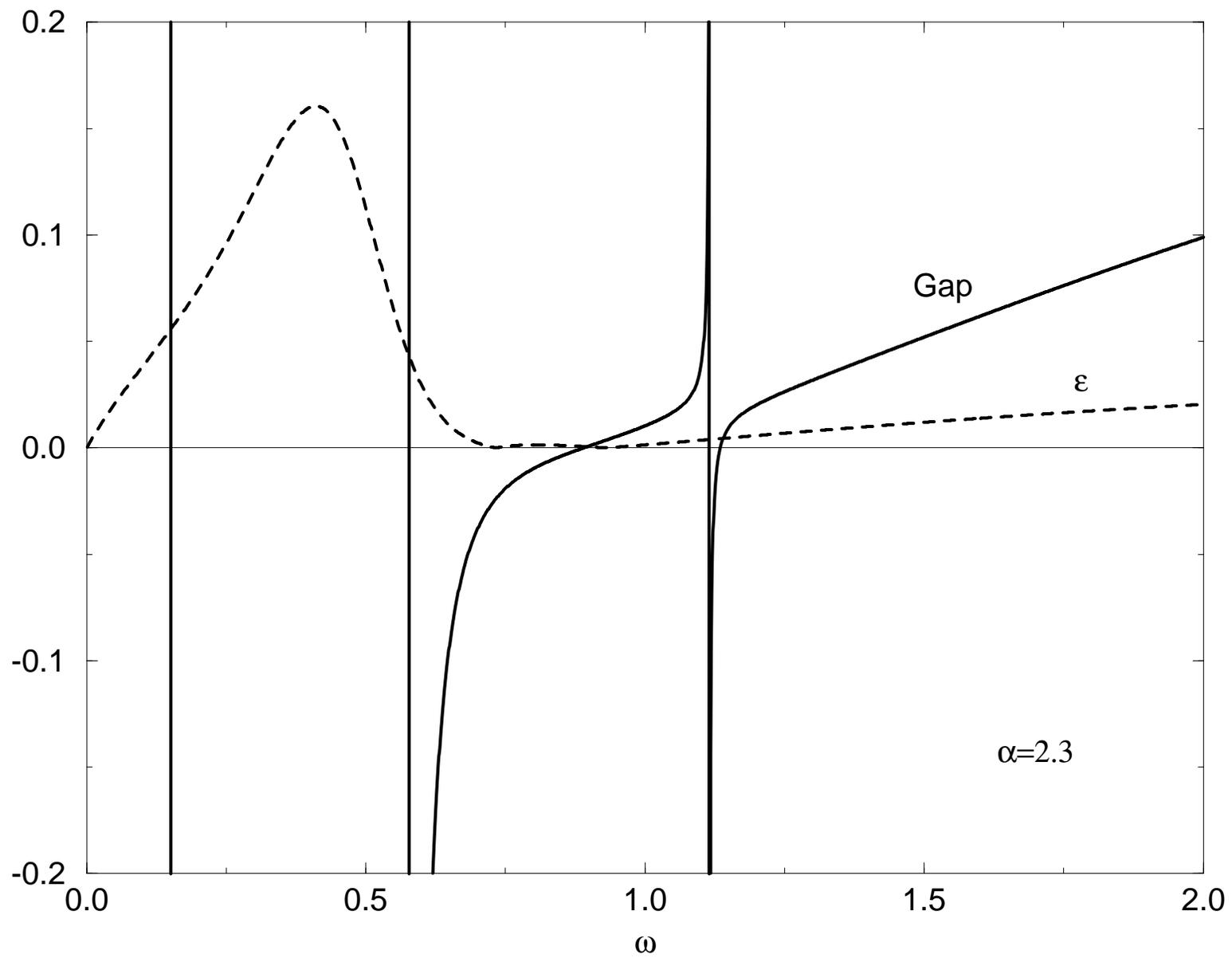

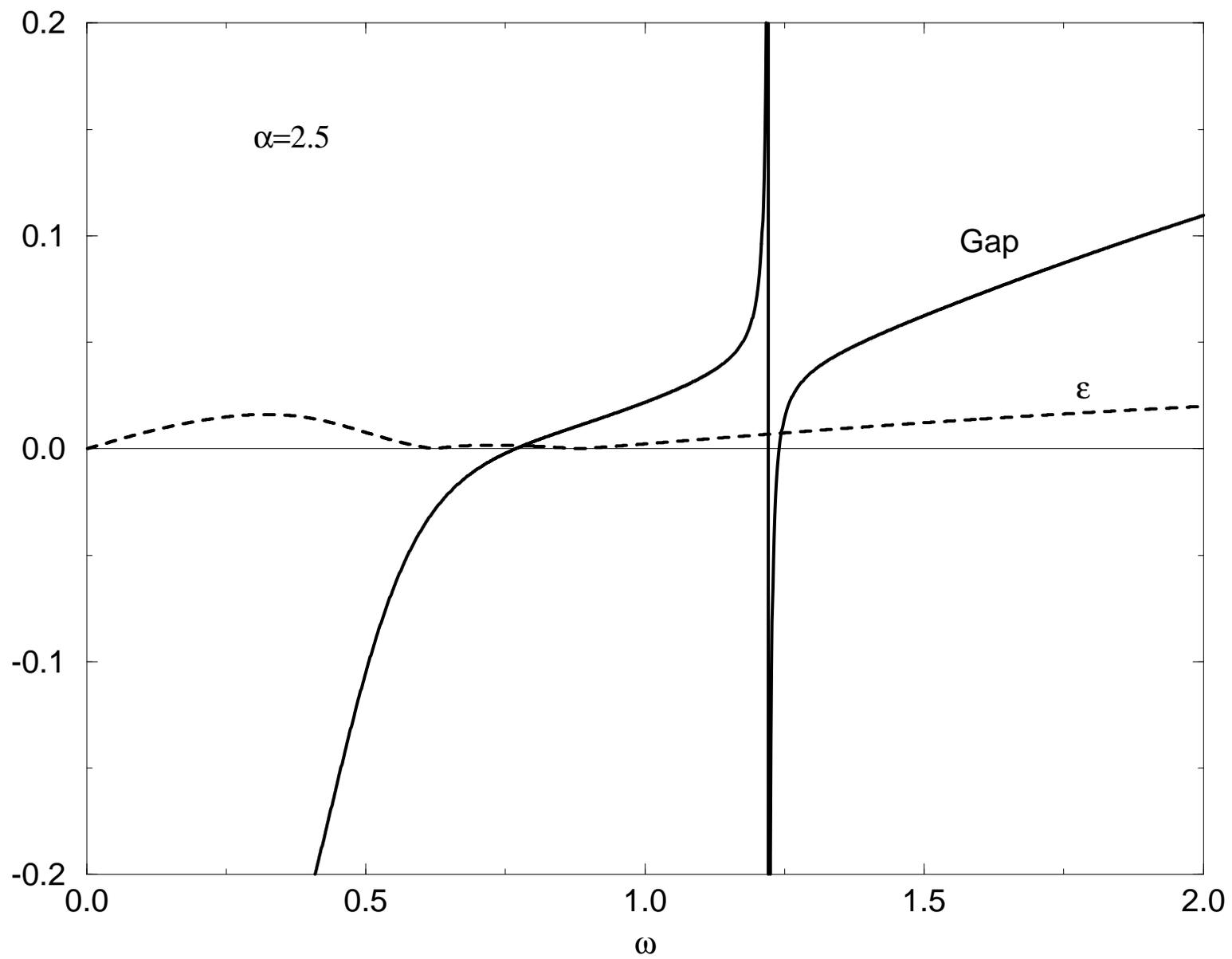

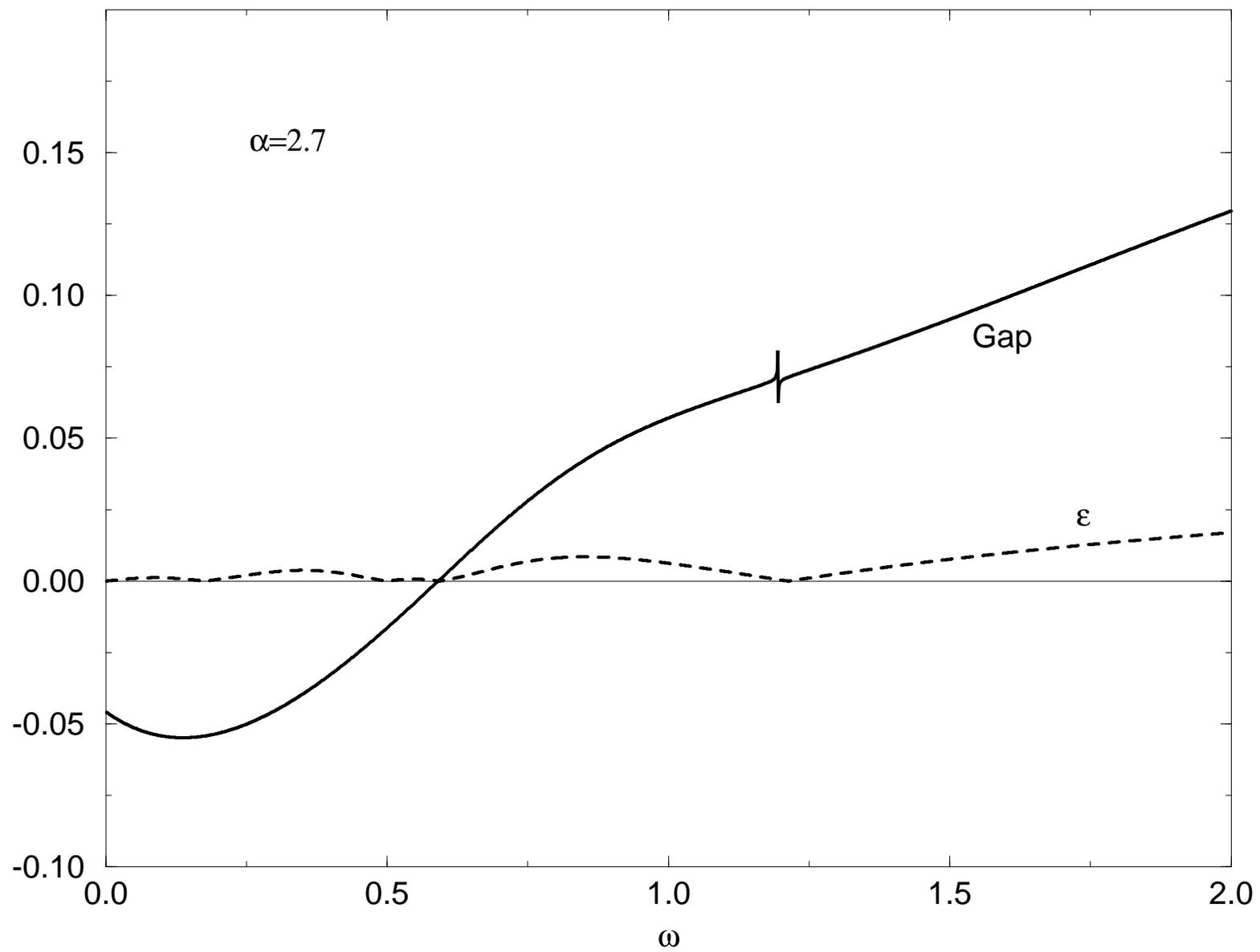

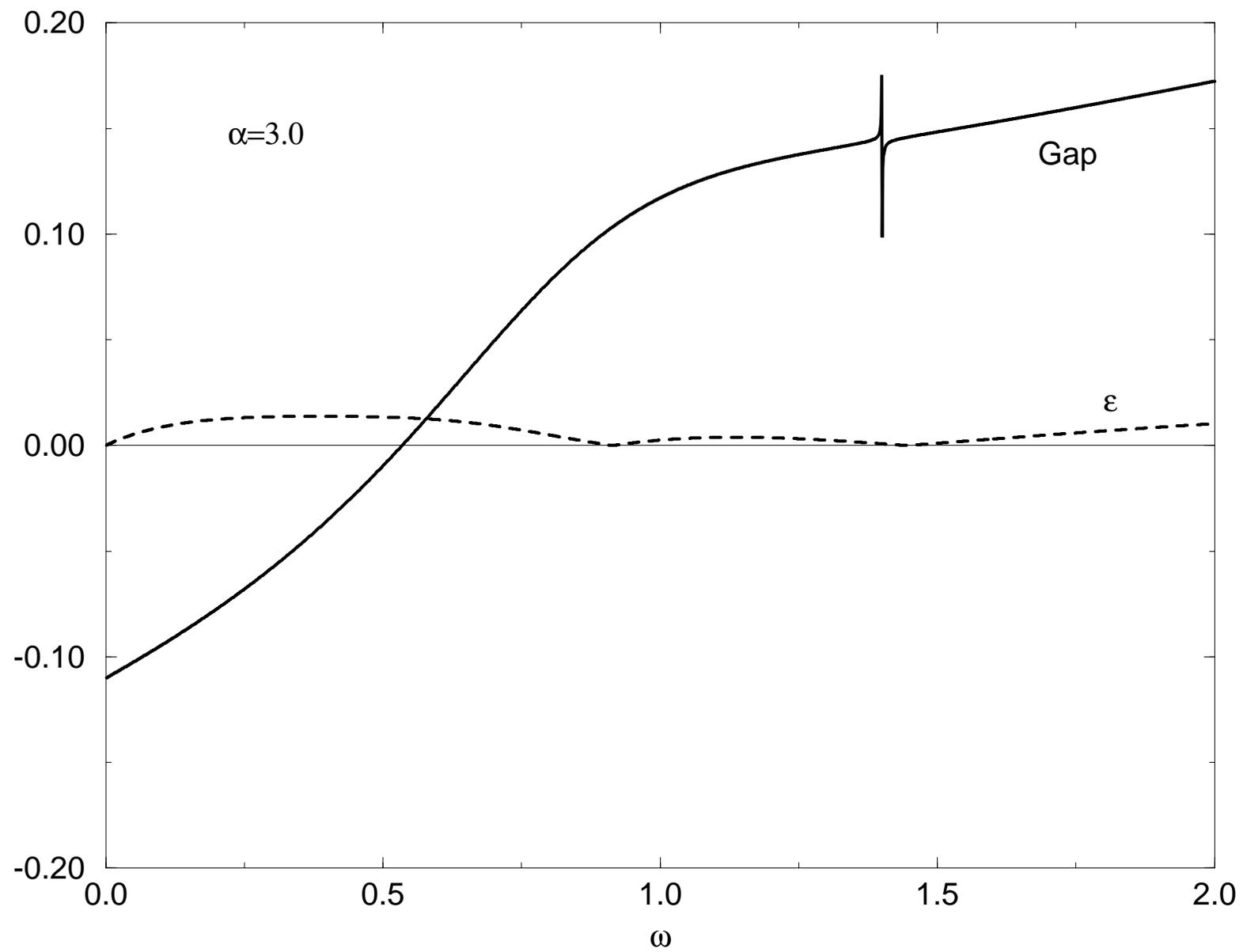

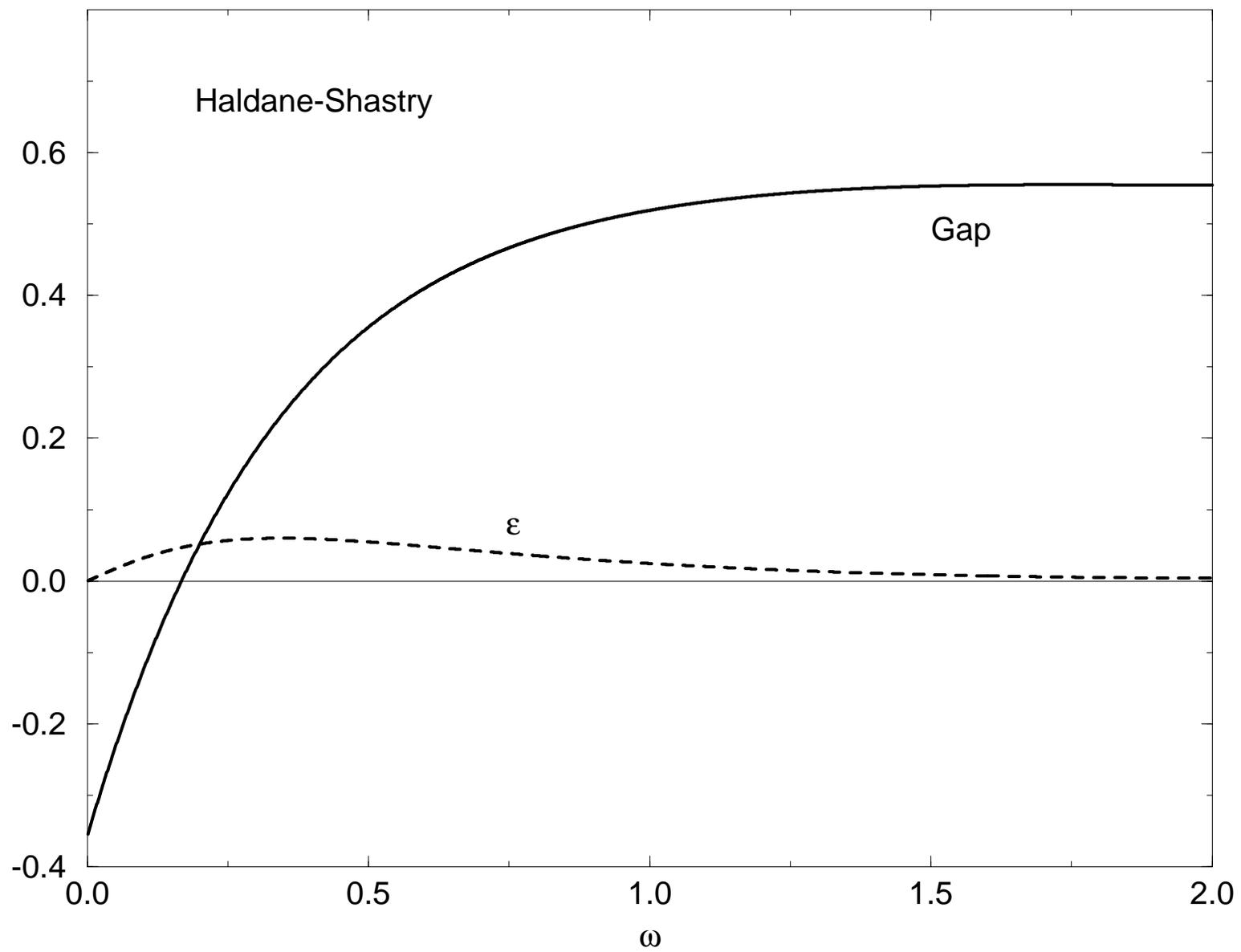

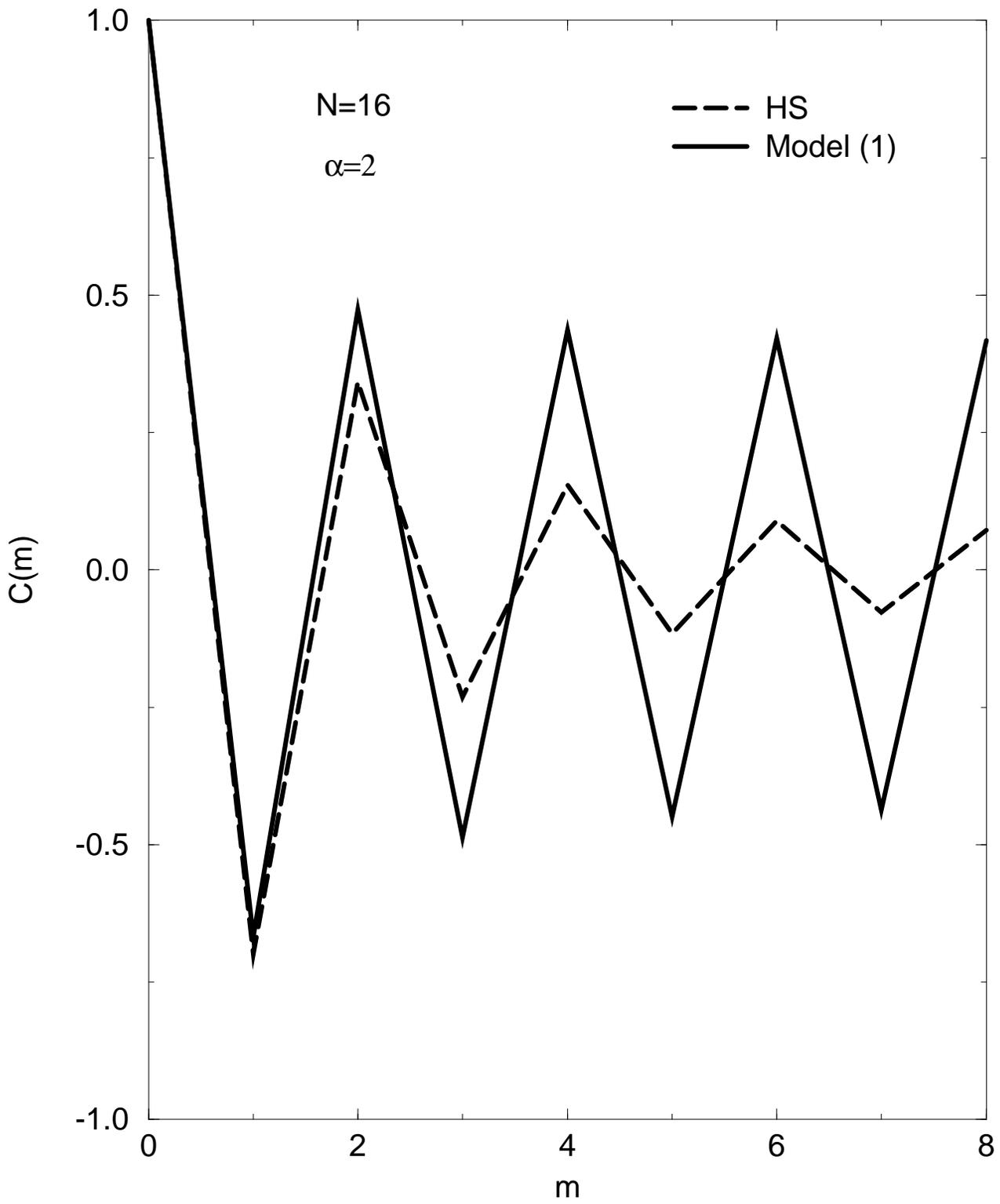

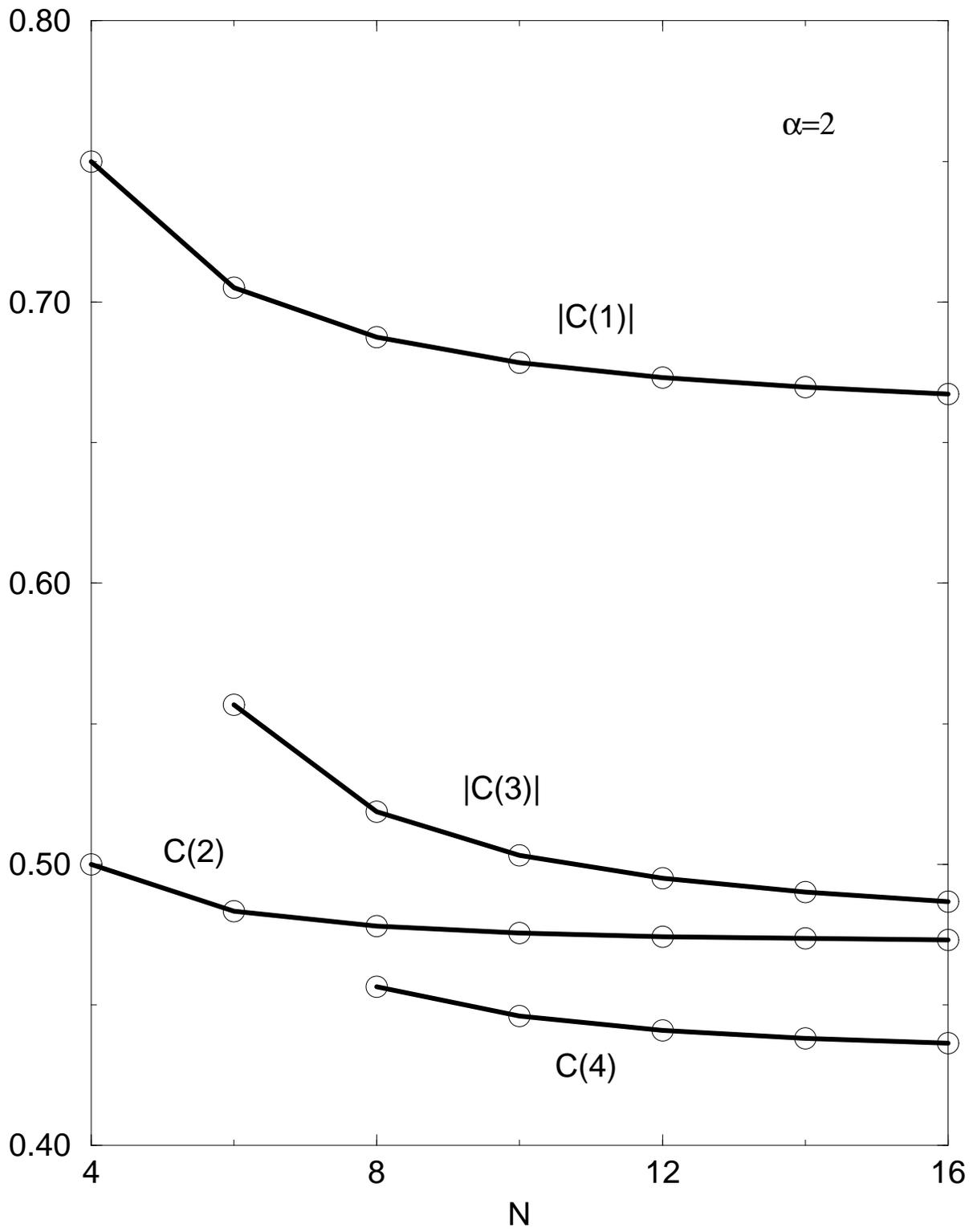

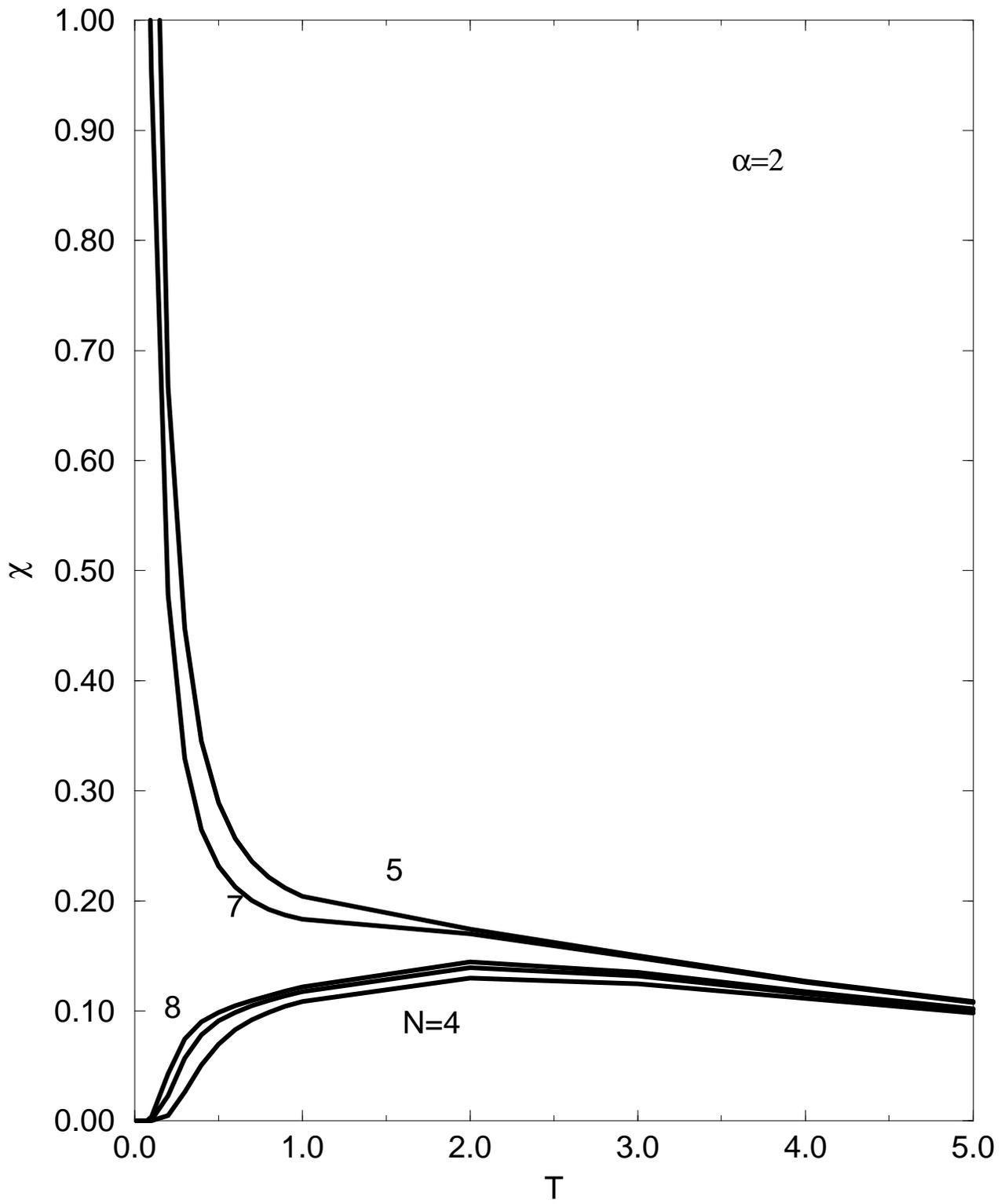

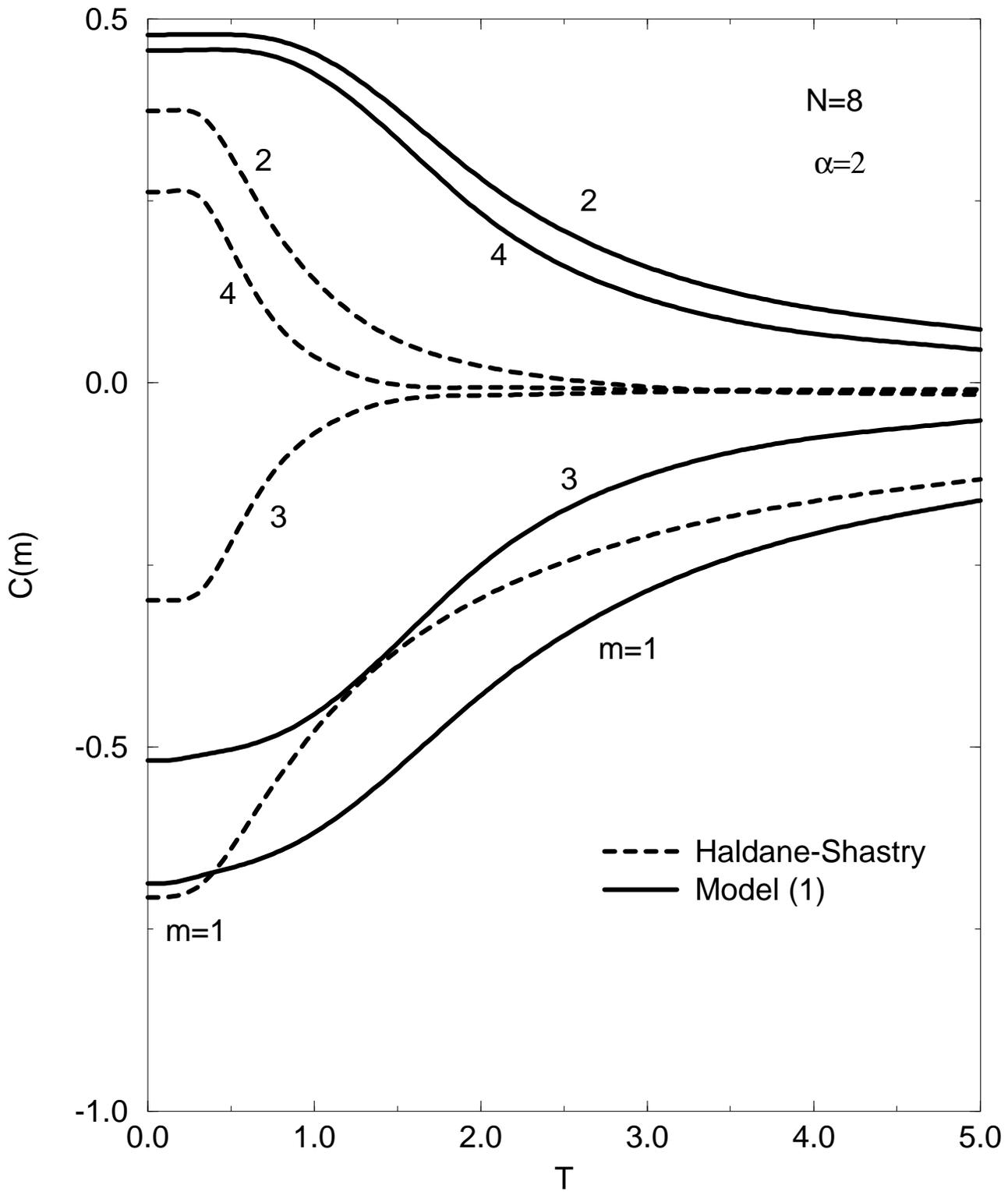

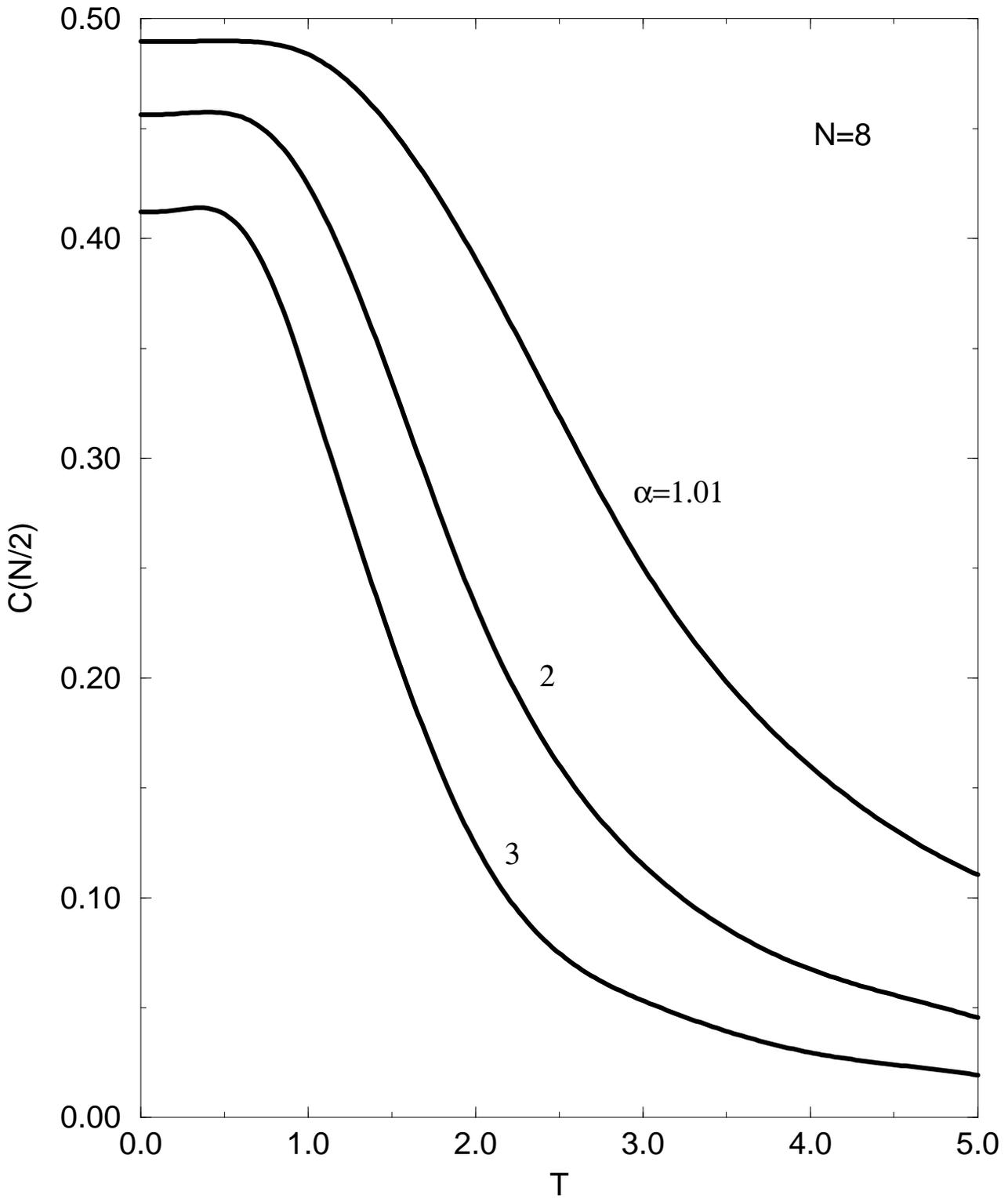

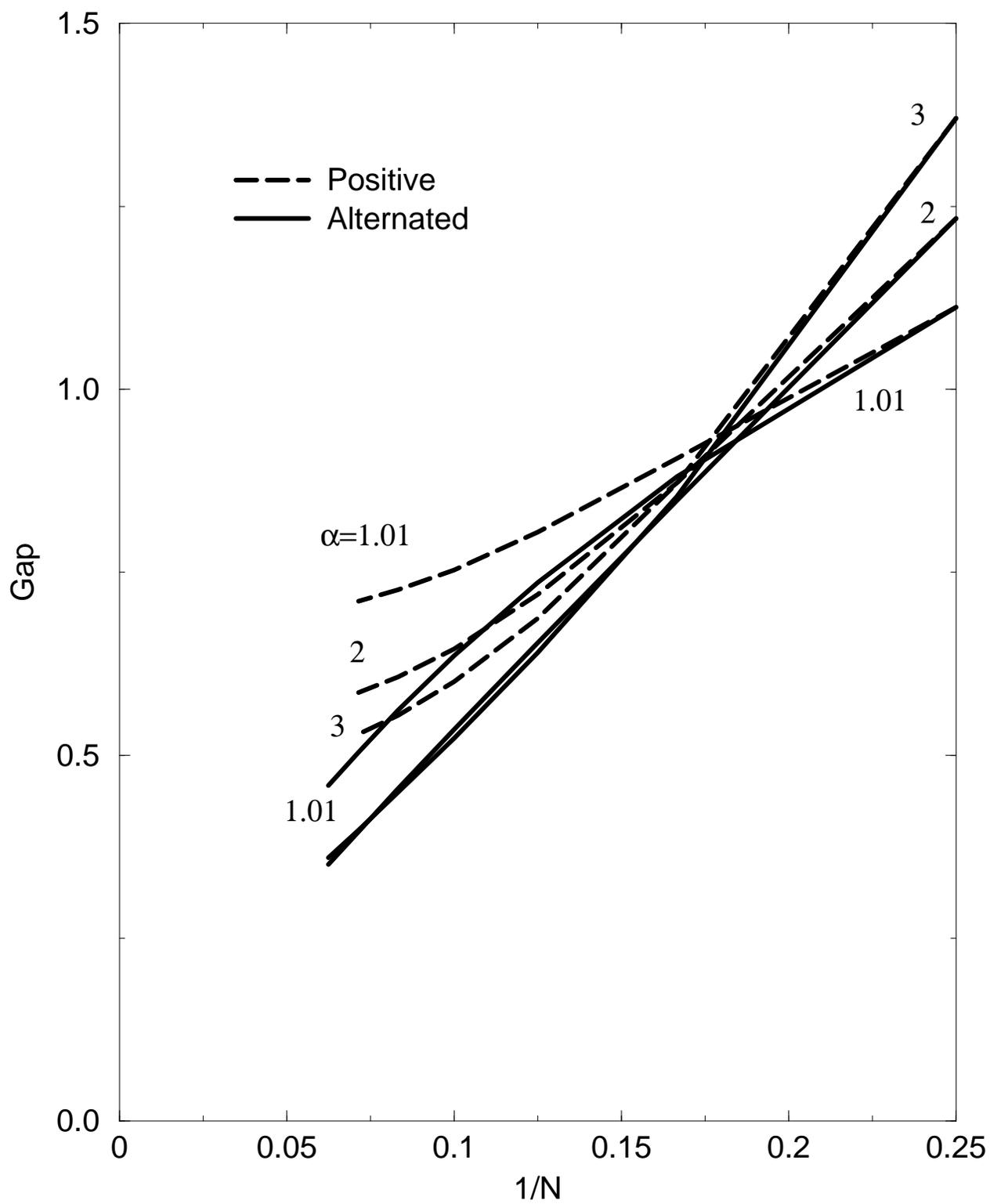

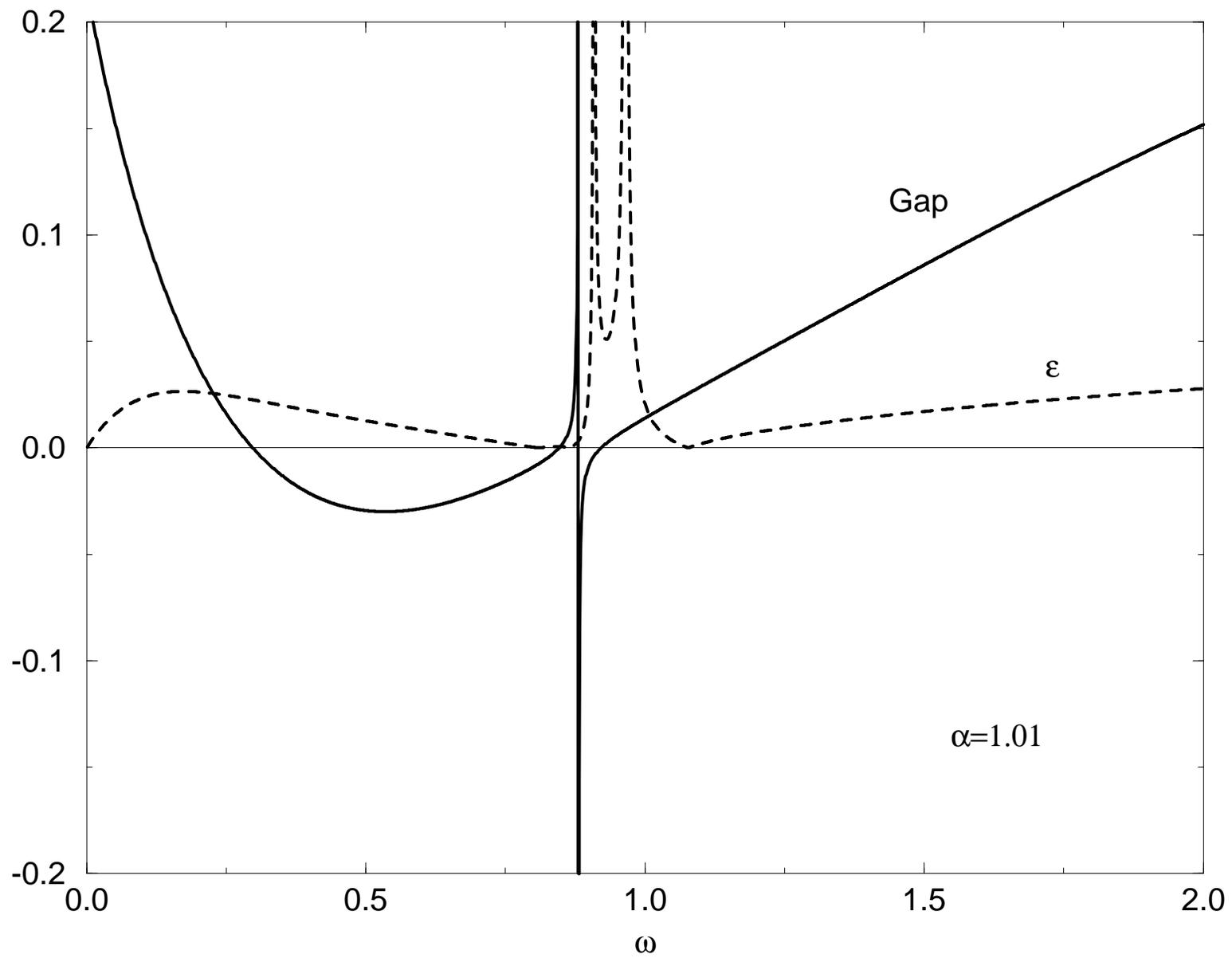

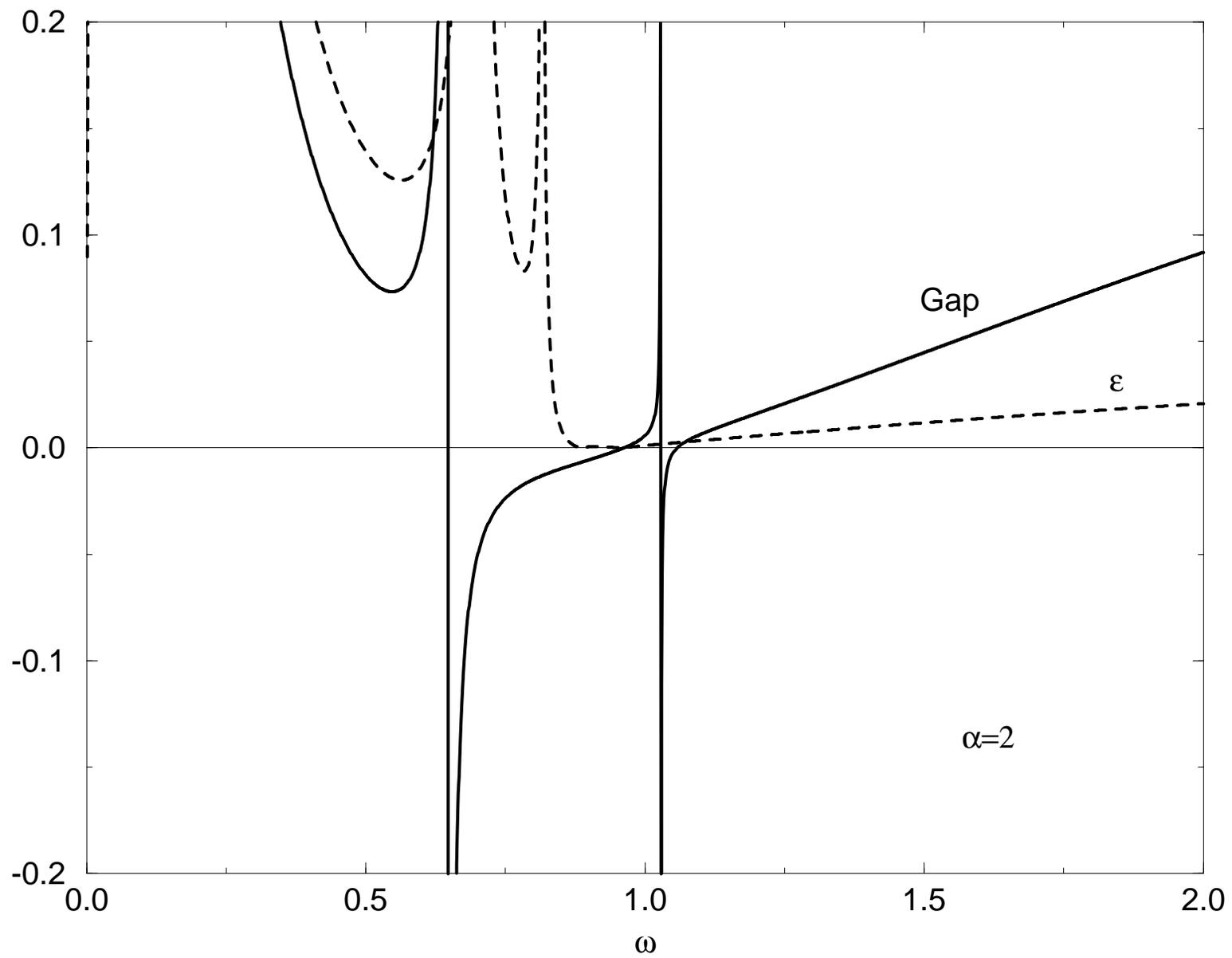

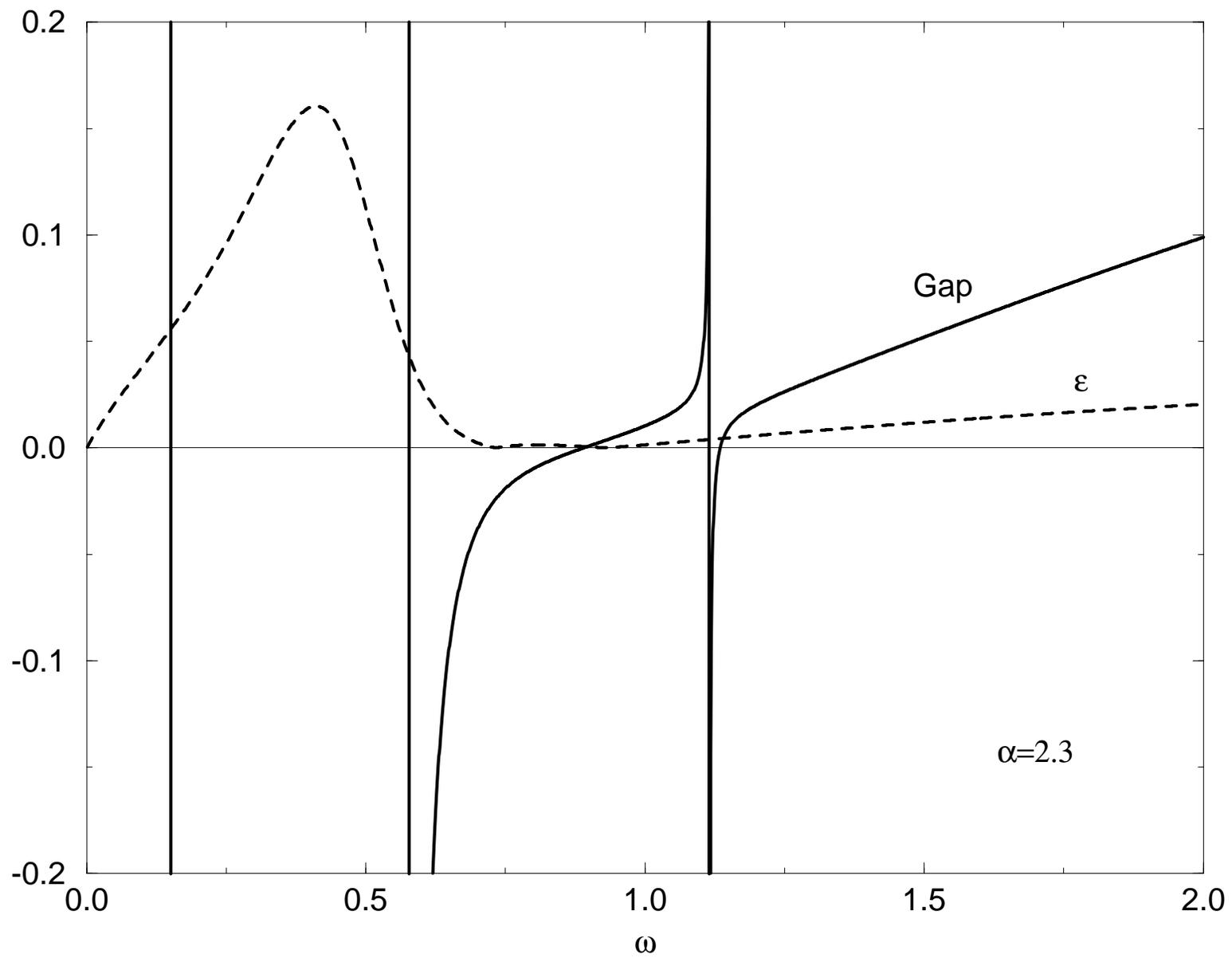

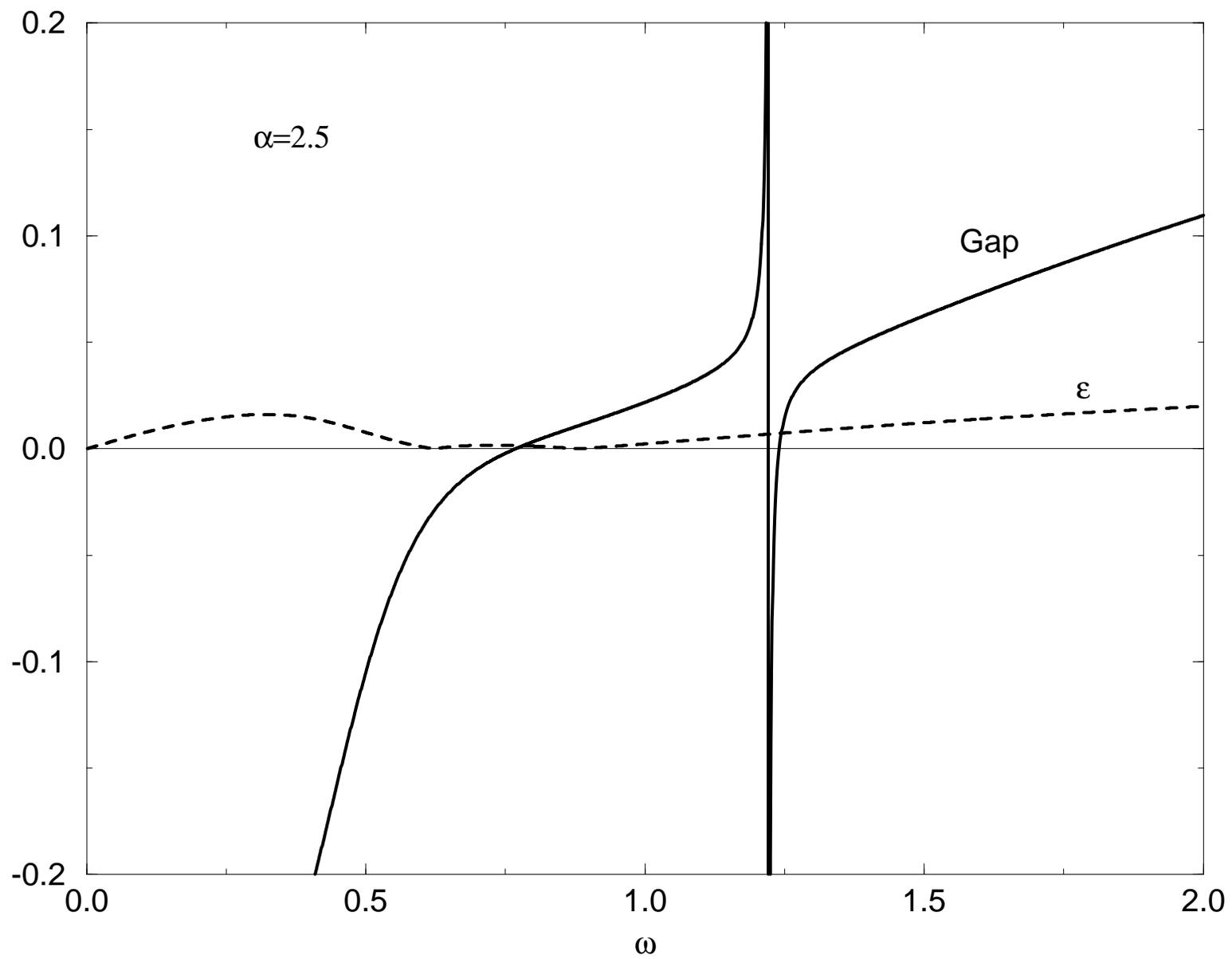

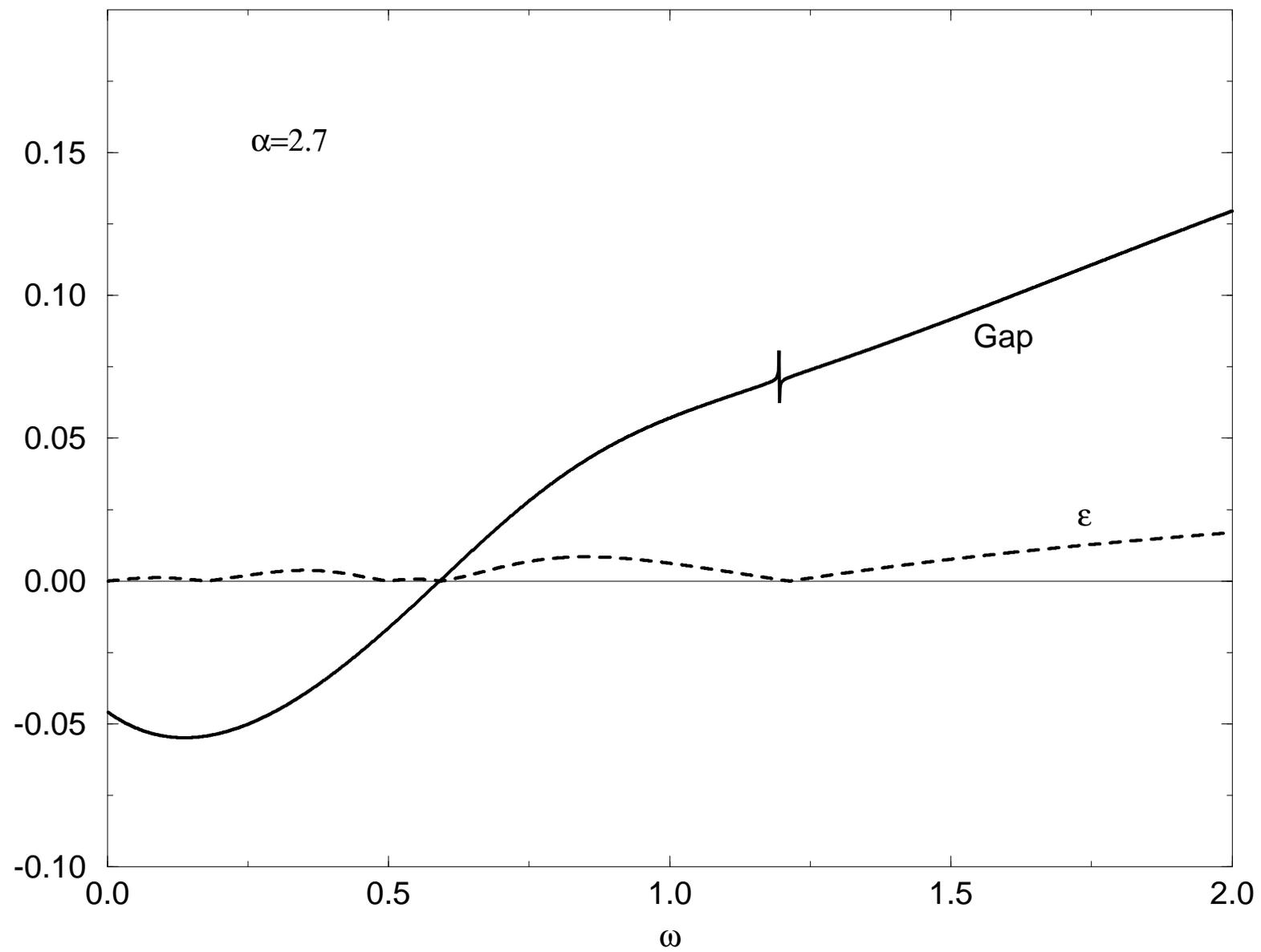

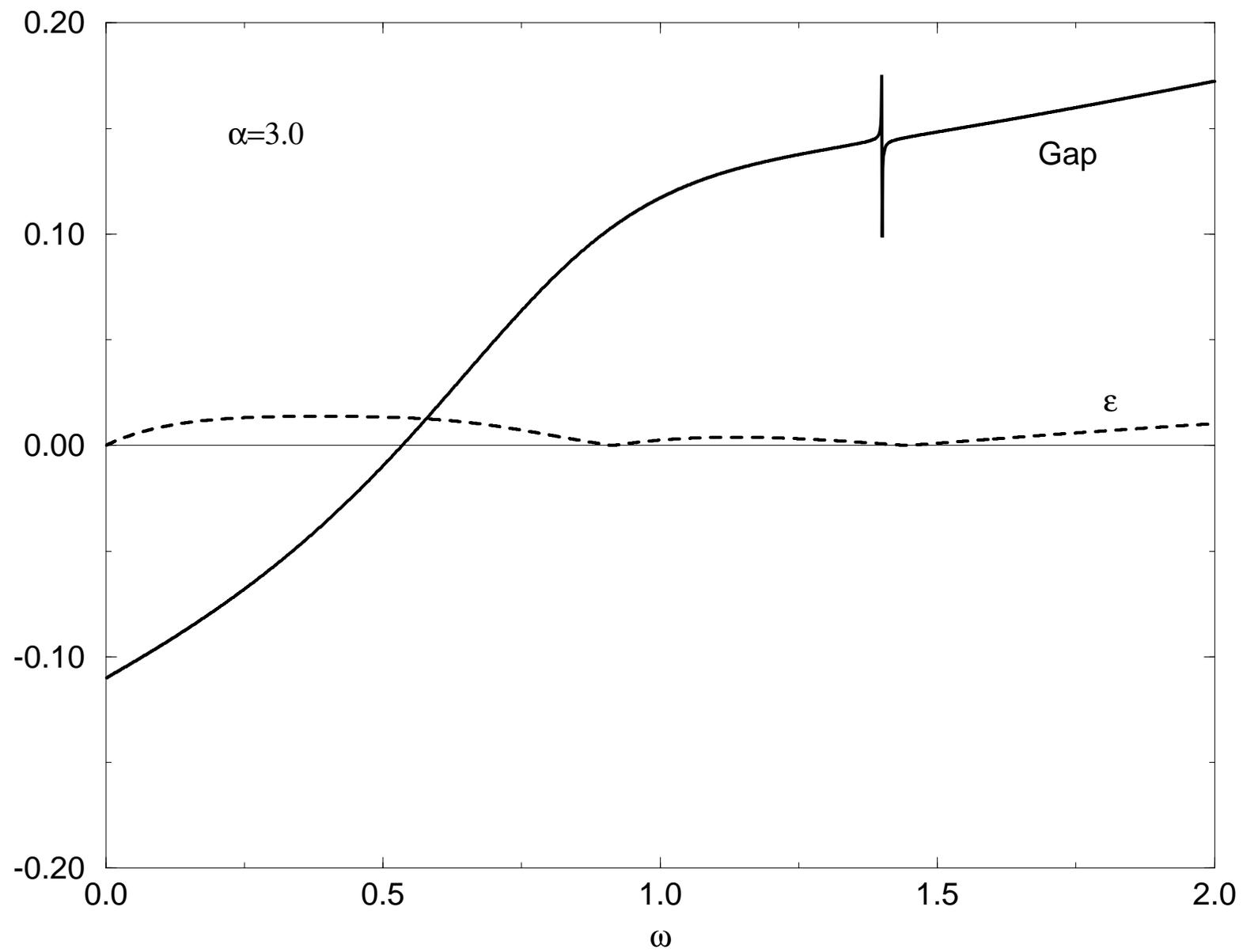

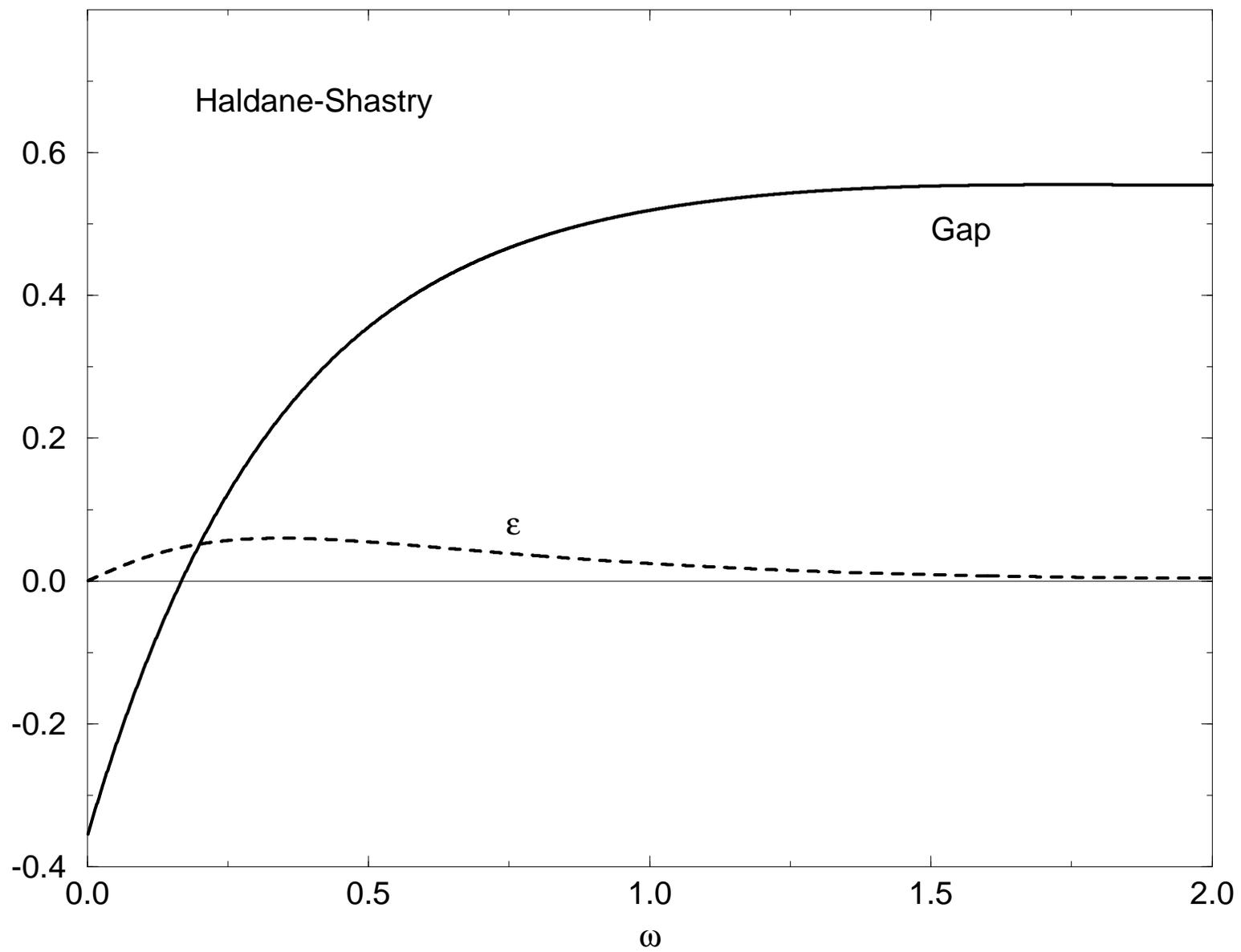

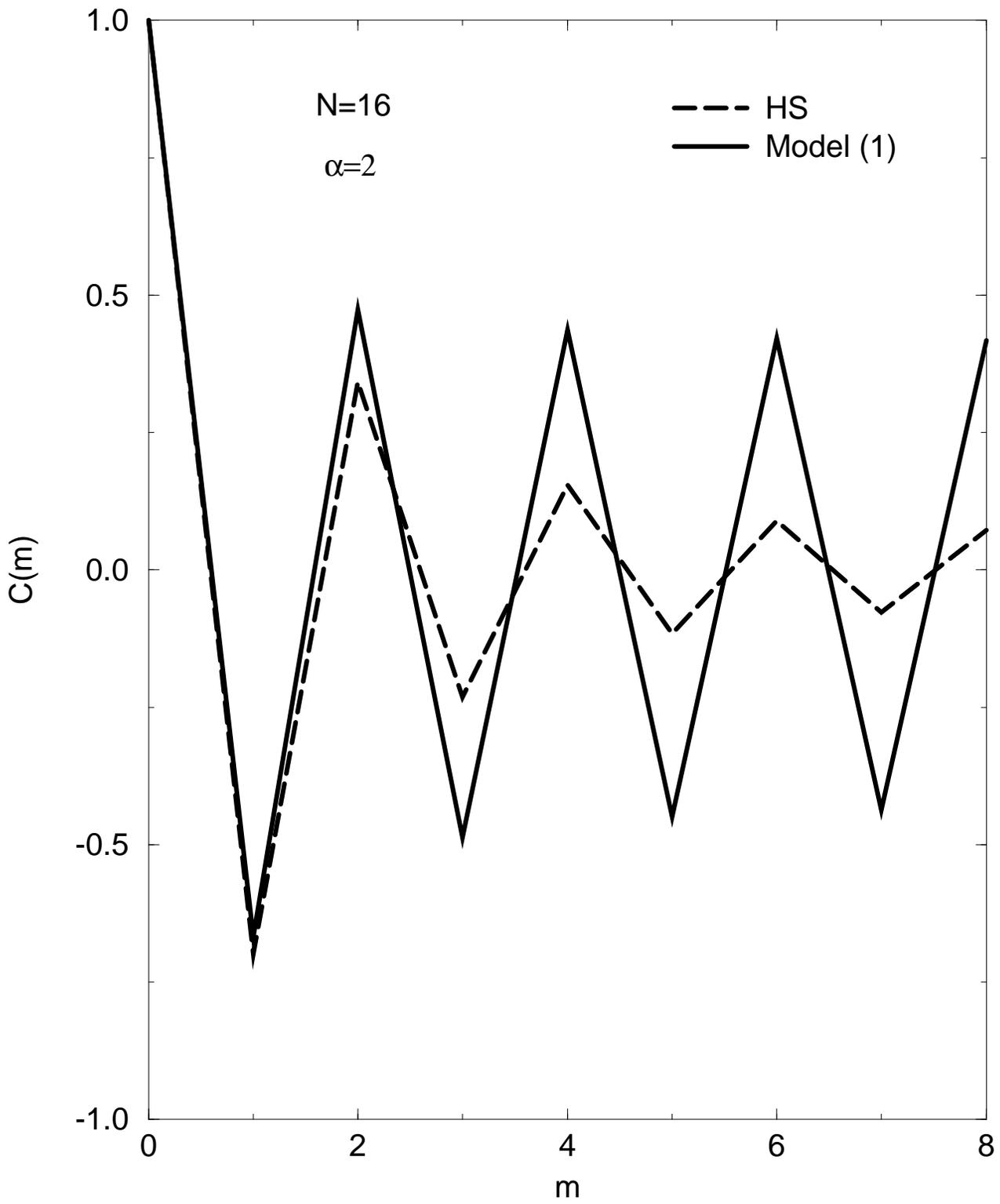

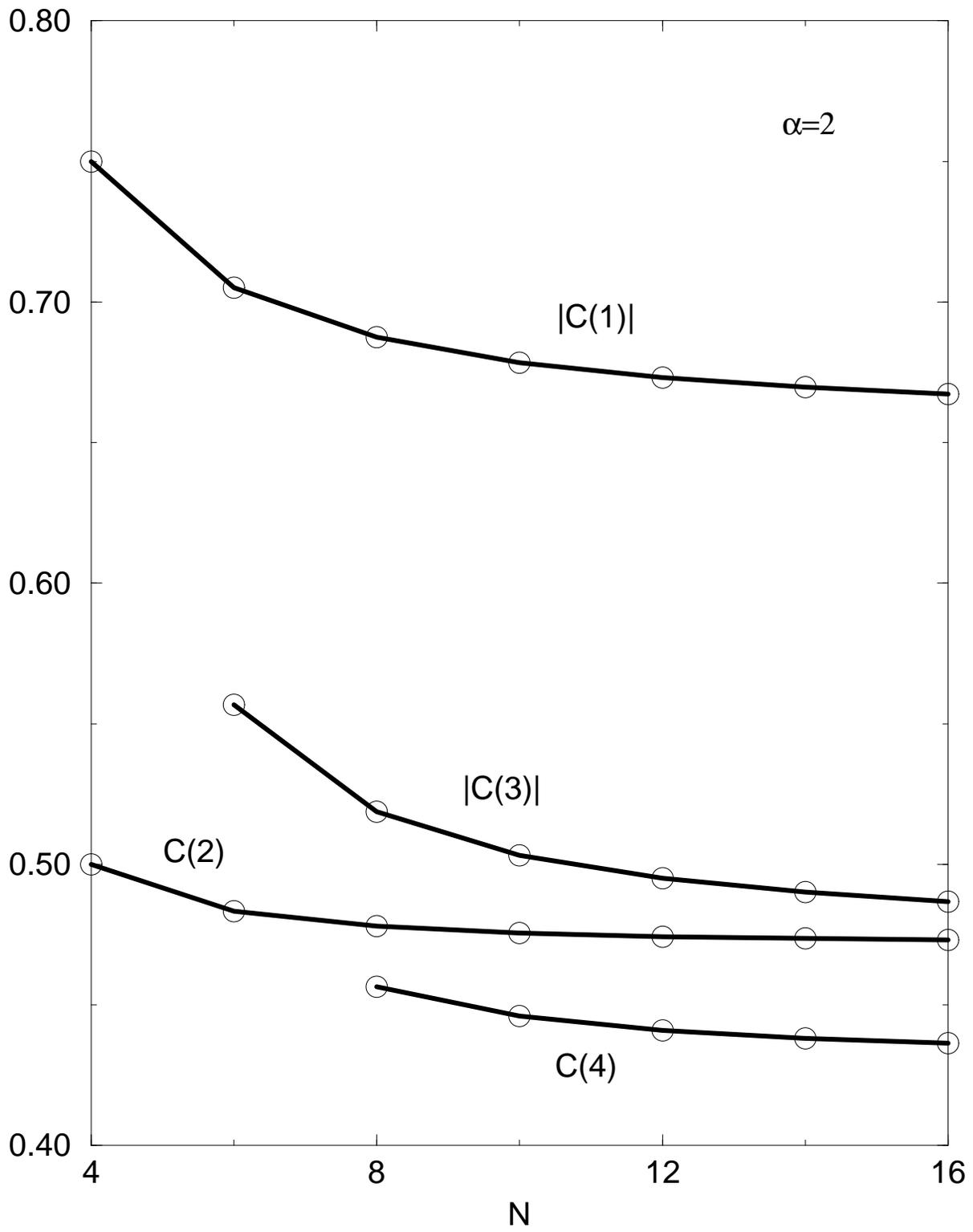

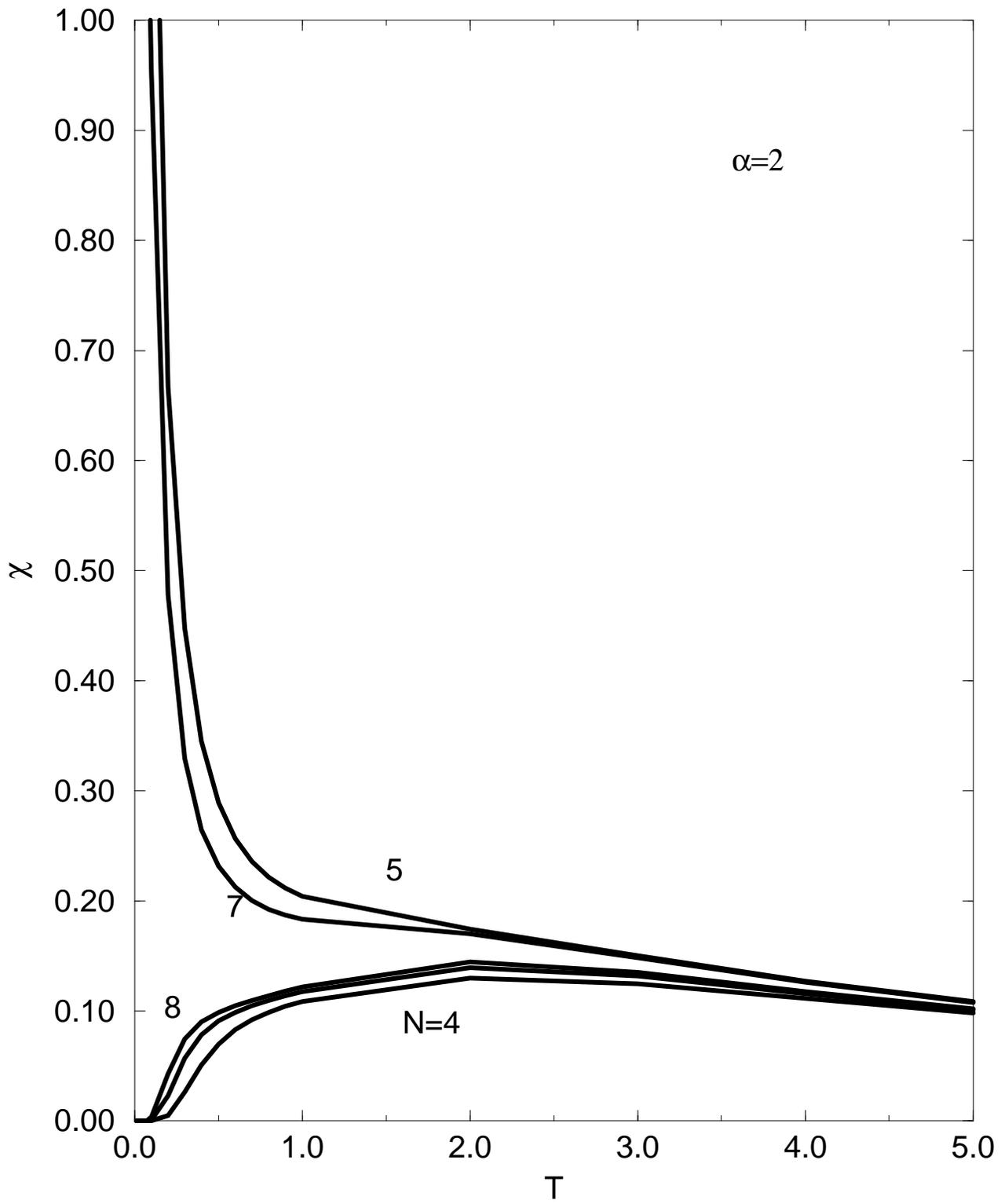

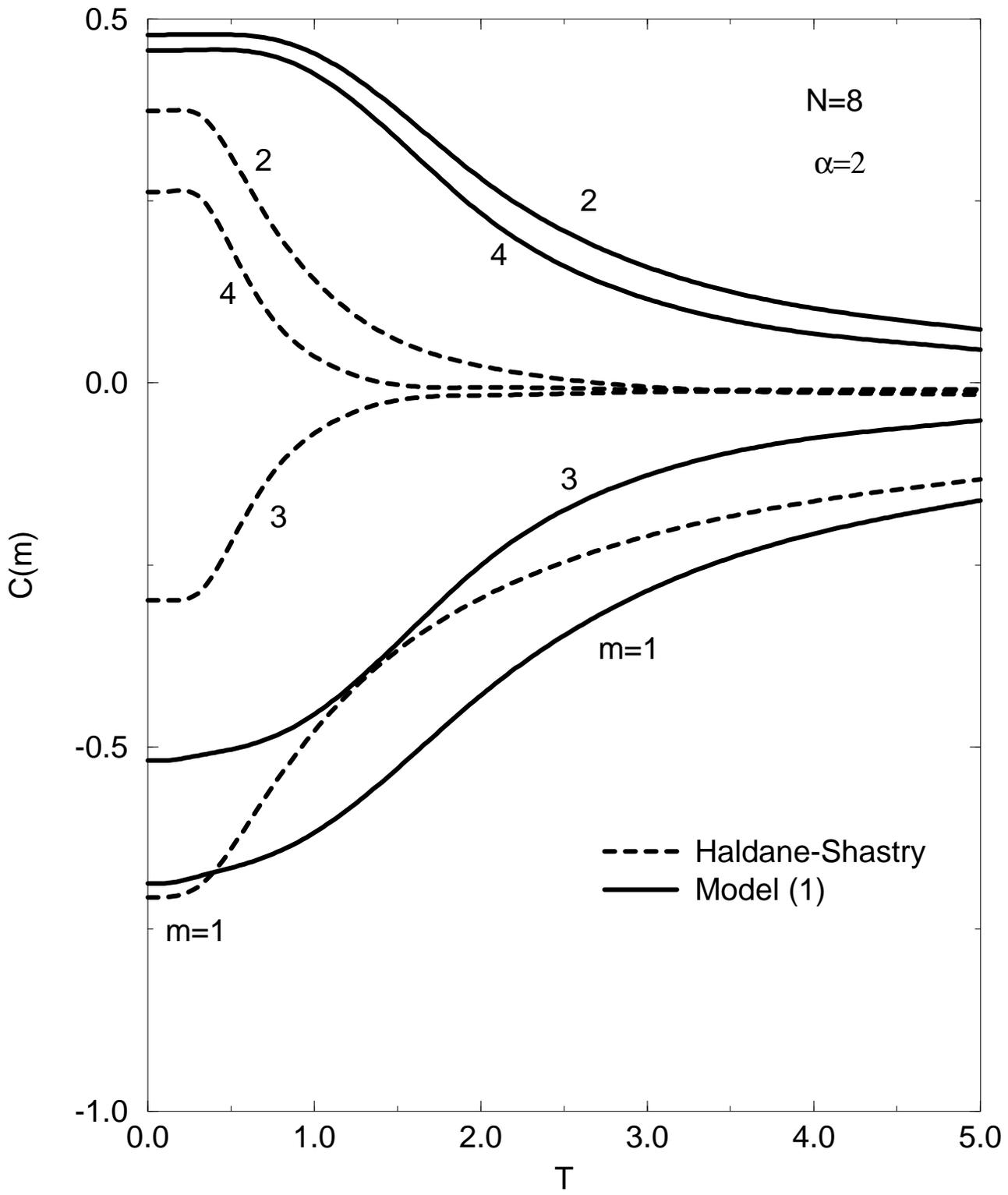

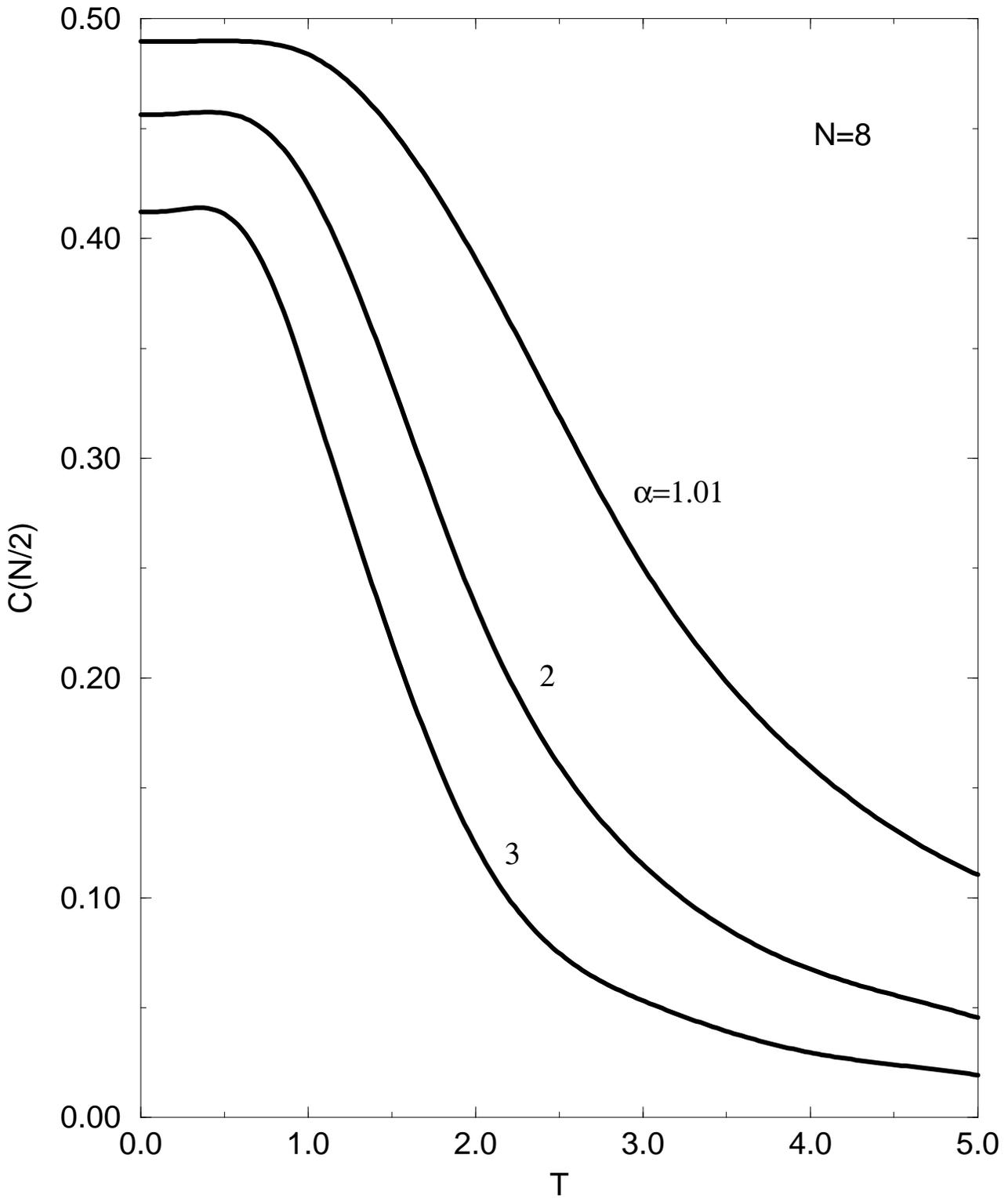

# Gapless spectrum in a class of $S = 1$ exchange models with long-range interactions

P.D. Sacramento[1,2] and V.R. Vieira[2]

[1] *Departamento de Física, Instituto Superior Técnico*

*Av. Rovisco Pais, 1096 Lisboa Codex, Portugal*

[2] *Centro de Física das Interacções Fundamentais, Instituto Superior Técnico*

*Av. Rovisco Pais, 1096 Lisboa Codex, Portugal*

We present evidence for the absence of a gap in a class of $S = 1$ antiferromagnetic exchange models. The spin exchange is long-ranged of the type $-(-1)^{i-j}/|i-j|^\alpha$ where $1 < \alpha < 3$. We have shown previously that without the alternating factor the model for $\alpha = 2$ ($S = 1$ Haldane-Shastry model) has a gap, exponentially decaying correlation functions and exponentially small susceptibility at very low temperatures. In the case of the alternated interaction the stabilizing next nearest neighbor ferromagnetic interaction changes qualitatively the behavior of the system. We have studied the groundstate and first excited state using a modified Lanczos algorithm for system sizes up to 16 sites. Also, we performed exact diagonalization for systems up to 8 sites and obtained the thermodynamics. The correlation functions decay with distance like a power law. These models define a new class of integer spin chains that do not show a Haldane gap. The results may be relevant to describe impurity spins coupled by a RKKY-interaction through a half-filled conduction-electron band.

Typeset Using *REVTEX*

*Introduction.* It was proposed long ago that integer and half-odd-integer spin chains behave qualitatively different [1]. The reason lies on a topological term in the action that prevails in the latter case. On general grounds half-odd-integer spin chains are gapless and integer spin chains show a gap [2]. This leads to important differences in the correlation functions. In the first case these show power law behavior and in the second case exponential behavior. There are however several models that do not follow this rule at special points in the space of interactions. In the case of $S = 1$, adding a quadratic term in the interaction and requiring the models to be integrable, a $SU(2)$ invariant model [3] and a $SU(3)$ invariant model (both solvable by the Bethe ansatz) [4] are gapless. Also, adding frustrating next-nearest-neighbor (nnn) interactions to the Heisenberg model, it has been found that for $S = 1/2$ there is a critical value $\alpha_{cr} \sim 0.2411$ (where $\alpha = J_2/J_1$, is the ratio of the nnn interaction to the nearest-neighbor (nn) interaction) such that for $\alpha < \alpha_{cr}$ the spectrum is gapless (as for $\alpha = 0$) while for $\alpha > \alpha_{cr}$ a gap appears [5]. This has been interpreted as a fluid-dimer transition. In particular, it has been shown that if the spin is half-odd-integer and the groundstate is translationally invariant ($k = 0$) the spectrum is gapless [6]. The dimer phase is consistent with the Lieb-Schultz-Mattis theorem [7] which states that if the spectrum is not gapless the groundstate should be degenerate (for half-integer spin). Also, recently it has been argued that translationally invariant spin chains in an applied field can be gapful without breaking translation symmetry, when the magnetization per spin, $m$, is such that $S - m$ is an integer. It was then proposed that a Haldane gap phase can be found for half-integer spin [8]. It has also been shown recently that the $S = 1/2$-dimer chain, the Majumdar-Ghosh chain and the $S = 1$-Haldane chain are in the same phase [9].

Explicit tests of the validity of Haldane's proposal have concentrated on models of short-range interactions. Recently, we have extended this analysis to the case of long-range interactions considering the $S = 1$ Haldane-Shastry model [10,11]. Even though we might expect that the correlation functions should not decay faster than the interaction we found a finite gap and the corresponding exponential decay of the correlation functions. This may be the result of the frustrating nature of the interactions. The influence of frustration in the $S = 1$

Heisenberg chain has also been studied recently [12].

The standard Haldane-Shastry model [13,14] is a periodic version of $1/r^2$ exchange. The $S = 1/2$ case has attracted considerable attention [15,16]. The groundstate energy and the correlation functions have been obtained [15,13,14] together with the thermodynamics [17]. The groundstate wavefunction is a spin singlet of the Jastrow-Gutzwiller form. The excitations are spin-1/2 spinons [17] that form a gas of a semionic nature [17,18]. The asymptotic correlations decay algebraically with exponent $\eta = 1$ without logarithmic corrections, in contrast to the Heisenberg case. This indicates the absence of spin exchange between the spinons rendering the models solvable in greater detail than in the short-range Heisenberg counterpart, solvable by the traditional Bethe ansatz method. The zero-$T$ susceptibility is finite [17] (and numerically the same as for the Heisenberg model) consistently with a singlet groundstate and a gapless spectrum.

The $S = 1$ case is not integrable (for both models). In ref. 10 we studied the groundstate properties and the gap to the first excited state using a modified Lanczos method for small systems of size up to 16 spins. We obtained that the groundstate energy per spin is $-1.267894$ and the value of the gap is 0.55439 (recall that for the Heisenberg model the gap has been estimated to be 0.41050). The general trend of the groundstate correlation functions is that they decay faster than those for the Heisenberg model (in the sense that the numerical values are smaller) both with distance for fixed $N$ and as a function of the size of the system. The same happens for the $S = 1/2$ case where the spectrum is gapless. A linear fit of $\log C_{N/2}$ as a function of $N$ yields a correlation length of the order of $\xi = 3.1$. This is an underestimate due to finite size effects. Also, we performed complete diagonalization of systems up to 8 sites to obtain the temperature dependence of the susceptibility and correlation functions [11]. The susceptibility is exponentially small at low-$T$ and the correlation functions also decay faster than those for the Heisenberg model as a function of temperature.

*Model.* In this work we consider a class of models of long-range interactions of the type $1/r^\alpha$ but with alternating signs [16]. A simple and convenient way to implement periodic boundary conditions is to use the chord distance (that is the distance between the points when the chain is wrapped into a closed circle). We consider then the Hamiltonian [19]

$$H = -\sum_{i<j}(-1)^{i-j}\frac{\vec{S}_i.\vec{S}_j}{[d(i-j)]^\alpha}. \qquad (1)$$

where $d(i-j)$ is the chord distance $d(i-j) = |\sin(i-j)\phi|/\phi$ with $\phi = \pi/N$, where $N$ is the number of lattice sites in the chain. The leading term (nn) is antiferromagnetic but the next nearest neighbor term (nnn) is ferromagnetic which tends to stabilize the dominant term. We take $1 < \alpha < 3$.

This class of long-range spin exchange models may describe impurity spins in a metallic host interacting via a RKKY-interaction $\sim \cos[2k_F r]/r^\alpha$ if the conduction electron band is half-filled. The value of the exponent $\alpha$ in the case of non-interacting electrons is $\alpha = d$ (where $d$ is the dimensionality of the lattice). The case of interacting electrons has also been considered. In the case of the magnetic screening cloud around a single Kondo impurity in a Luttinger liquid ($d = 1$) $\alpha = g_c < 1$ [20] and in the case of the coupling between two impurity spins ($S = 1/2$) coupled to a $1d$ Hubbard chain $\alpha = 2$ [21], for example.

To study Model (1) we use the modified Lanczos algorithm [22] to obtain the groundstate properties and the gap to the first excited state and we use complete diagonalization of the Hamiltonian matrix to obtain the thermodynamics. In the modified Lanczos algorithm the size of the vectors can be considerably reduced using the symmetries of the problem. The Hamiltonian commutes with the total spin operators $(\vec{S}_T)^2$ and $S_T^z$, with the translation operator $T$, the spin flip operator $R$ and the reflection operator $L$ ($i \to N + 1 - i, i = 1, \ldots, N$). The ground state has total $S_T^z = 0$ and one of the (degenerate) first excited states also has $S_T^z = 0$. One can then immediately reduce the states under consideration to this subspace only [10]. A similar procedure can be used for the finite temperature behavior. We calculate the susceptibility

$$\chi = \frac{\beta}{3N}\sum_{i,j}<\vec{S}_i.\vec{S}_j> \qquad (2)$$

and the correlation functions

$$C_m = \frac{3}{N}\sum_{i=1}^{N}\frac{<S_i^Z S_{i+m}^z>}{S(S+1)}. \qquad (3)$$

Both diagonalizations give the exact results for the several finite-size systems. The results for the infinite system can be estimated using standard extrapolation methods [23] like the VBS method [24] or the BST method [25]. In the first method we want to estimate the limit of a finite sequence $P_n$ ($n = 1, ..., N$). Defining

$$P_n^{(m+1)} = P_n^{(m)} + \frac{1}{Q_n^{(m)} - Q_{n-1}^{(m)}} \quad (4)$$

$$Q_n^{(m)} = \alpha_m Q_n^{(m-1)} + \frac{1}{P_{n+1}^{(m)} - P_n^{(m)}} \quad (5)$$

where $Q_n^{(-1)} = 0$, $P_n^{(0)} = P_n$ we obtain an estimate of the sequence iterating. If $\alpha_m = 0$ this is the Aitken-Shanks transformation which is adequate for exponential behavior. To generate the Padé-Shanks transformation we select $\alpha_m = 1$. A power law behavior is well fitted choosing the Hamer and Barber's transformation $\alpha_m = -[1 - (-1)^m]/2$. We get an estimate of the asymptotic value of the sequence $P_n$ [23,24] selecting $\alpha_m$ appropriately.

In the BST algorithm we look for the limit of a sequence of the type $T(h) = T + a_1 h^\omega + a_2 h^{2\omega} + ...$, where $h_N = 1/N$ is a sequence for the several system sizes, $N$. The value of the $m^{th}$ iteration for the sequence is obtained from

$$T_m^{(N)} = T_{m-1}^{(N+1)} + \frac{T_{m-1}^{(N+1)} - T_{m-1}^{(N)}}{\left(\frac{h_N}{h_{N+m}}\right)^\omega \left(1 - \frac{T_{m-1}^{(N+1)} - T_{m-1}^{(N)}}{T_{m-1}^{(N+1)} - T_{m-2}^{(N+1)}}\right) - 1} \quad (6)$$

where $\omega$ is a free parameter which is adjusted such that the estimate of the error

$$\epsilon = |T_{N_p-2}^{(1)} - T_{N_p-2}^{(0)}| \quad (7)$$

is a minimum (where $N_p$ is the number of data points, $T_{-1}^{(N)} = 0$ and $T_0^{(N)} = T(h_N)$). It has been shown [25] that this algorithm has several advantages over the VBS algorithm in particular for smaller sequences, it converges faster and it is less sensitive to rounding errors. In this paper we will use both methods to estimate the thermodynamic limit.

*Results.* In Table I we show the values for the groundstate energy and gap as a function of $N$ (for $N = 4$ to $16$ and $N$ even) for the values of $\alpha = 1.01, 2, 3$ [26]. In Fig. 1 we show the gap as a function of $1/N$ for the standard frustrating case compared to the behavior of the gap in the alternating case. The results suggest that these two cases are in different classes since the slopes for the two cases are distinct. In particular, it suggests that the alternated cases may extrapolate to a zero value of the gap as $N \rightarrow \infty$ while the standard frustrating cases suggest a finite value in agreement with the results previously obtained for $\alpha = 2$ [10]. In Table II we show the VBS estimated values for the gap using the several extrapolations. For small and intermediate values of $\alpha$ the power law fit (Hamer and Barber) consistently gives very small gaps suggesting that the spectrum is gapless. As $\alpha$ grows further ($\alpha \geq 2.7$) we see that the gap increases and becomes apparently finite (we will return to this point later). The table shows consistently smaller gaps for the power law fit as compared to the other two methods. If the spectrum is indeed gapless this is consistent since, for instance for the Aitken-Shanks transformation, one is trying to fit a power law with an exponential. In Tables III-V we present the sequence of $P_n^{(m)}$ using the Hamer-Barber algorithm for $\alpha = 2, 2.5, 3$. The difference between the extrapolated value and the values for the previous iteration gives a measure for the error involved. Due to the small number of data points this difference is actually an order of magnitude larger than the extrapolated value itself (except for $\alpha = 3$). Another possible criterium is to take the difference between the values for $m = 2$, but is also of the same order. The error in this procedure is therefore large.

To further clarify the nature of the spectrum and to have a better control on the errors involved we consider now the BST algorithm eq. (6). In Fig. 2 we present the values of the gap as a function of the free parameter $\omega$ for the several values of $\alpha$ considered above. We also plot the error $\epsilon$ defined in eq. (7). In general, for each value of $\alpha$ there are several points in $\omega$-space where either the gap or the error go to zero or become very small. For each value of $\alpha$ there is a point (actually a narrow region) where *both* are very small. First we take the value of $\omega$ where the gap is very small and estimate its error calculating $\epsilon$ at that point. In Table VI we give the error $\epsilon$ at the values of $\omega$ chosen as above for the several values of $\alpha$. We can also follow the standard method and select the several local minima for $\epsilon$ and take the gap obtained at these points. The results are shown in Table VII. They strongly suggest that the spectrum is gapless (at least for $\alpha \leq 2.7$). The BST method is consistent with the VBS method for $\alpha = 3$ in the sense that it suggests a finite gap (see however ahead).

Even though the results strongly suggest that the spectrum is gapless, it might appear that Table II could be consistent with a finite gap (but small compared to the frustrated case [10]). However, if the gap would be finite the Hamer-Barber algorithm should yield a finite gap as we obtained previously for the frustrated case [10] where all three algorithms correctly picked up the leading (finite) term of the sequence (see Table 4 of ref. 10). As a further check we have used the BST algorithm to find the extrapolated gap for the standard Haldane-Shastry model ($\alpha = 2$, frustrated case). The results are shown in Fig. 3. As a function of $\omega$ the error decreases for $\omega \sim 2$ and the gap is saturated to the value $0.55405$ (for $\omega = 2$) very close to the (finite) result previously obtained [10]. When there is a true gap the extrapolation yields a finite value with a magnitude that is very close for the *four* algorithms (actually one might take the discrepancy of extrapolations of Table II as a sign that the gap is zero - no finite leading term in the sequence).

In Table VIII we present the groundstate correlation functions for $\alpha = 2$. In Fig. 4a we show the groundstate correlation functions for $\alpha = 2$ and $N = 16$ for both cases showing that the decay is considerably slower in the alternated case. In Fig. 4b we show the correlation functions $C(m)$ for $m = 1-4$ as a function of $N$ for $\alpha = 2$. The behavior of $|C(1)|$ is similar for the positive and alternated interactions (it is slightly smaller in the latter case). However, for $m > 1$ the correlation functions $C_{alt}(m)$ are considerably larger and reach their extrapolated values for much smaller system sizes. We estimate the correlation function exponent taking $C(N/2) \sim (-1)^{N/2} 1/N^\eta$ [8] since for the system sizes considered it yields better results than a plot of $C(m)$ as a function of $m$ for fixed $N$ [8]. We estimate $\eta = 0.16, 0.24, 0.49$, for $\alpha = 1.01, 2, 3$, respectively (taking a fit using the system sizes up to $N = 16$ and excluding the $N = 4$ point). Besides the alternating signal, the correlation functions are modulated by an oscillatory function that decreases in amplitude as $N$ grows. Note that $|C_{alt}(2m+1)| > |C_{alt}(2m)|$ for $m \geq 1$ due to the oscillatory nature of the interaction (in Fig. 4b this is explicitly shown for $|C(3)| > C(2)$).

In Fig. 5 we show the susceptibility as a function of temperature for $\alpha = 2$. For $N$

even the groundstate is a singlet but for $N$ odd is degenerate (triplet) (as for the $S = 1/2$ Heisenberg model). Therefore $\chi$ alternates between zero and very large at small $T$. Comparing with results obtained for other models (like the Heisenberg and the Haldane-Shastry models for integer and half-integer spins) this alternancy suggests a gapless spectrum (in the case of a gap both the even and odd system sizes give a vanishing susceptibility at zero temperature). This would imply a finite value for the susceptibility (the extrapolation error is large and we do not estimate the zero-$T$ susceptibility). As $\alpha \to 3$ we find the same type of behavior indicating that there is no true gap and that the extrapolated results are a consequence of the finiteness of the systems studied. The reason is that the decay of the interaction is faster (and therefore the transition to gapless behavior is slower) and the finite sizes considered are not enough to correctly extrapolate to zero.

In Fig. 6 we show the correlation functions as a function of $T$ for the positive case and the alternating case for $\alpha = 2$. Consistently with the groundstate results the correlation functions also decay much slower with temperature in the alternated case. In Fig. 7 we show $C(N/2)$ for $N = 8$ for the several values of $\alpha$.

In summary, we have identified a new class of integer spin chains that are gapless. The interaction is long-ranged and non-frustrating. Indeed, the stabilizing ferromagnetic next-nearest-neighbor interaction changes qualitatively the behavior of the system with respect to an antiferromagnetic frustrating nnn interaction [10,11]. We calculated the gap (for $S = 1$) using a modified Lanczos method finding a vanishing value in the extrapolated limit. We also found that the correlation functions have a considerably larger range both with distance and as a function of temperature. The results obtained may be useful to study $S = 1$ impurity spins embedded in a half-filled conduction electron matrix and coupled via a RKKY-interaction.

# REFERENCES

[1] F.D.M. Haldane, Phys. Lett. **93**A, 464 (1983); Phys. Rev. Lett. **50**, 1153 (1983).

[2] I. Affleck, T. Kennedy, E.H. Lieb and H. Tasaki, Phys. Rev. Lett. **59**, 799 (1987); Comm. Math. Phys. **115**, 477 (1988).

[3] L. Takhtajan, Phys. Lett. **87**A, 479 (1982); H.M. Babujian, Phys. Lett. **90**A, 479 (1982); Nucl. Phys. B **215**, 317 (1983).

[4] B. Sutherland, Phys. Rev. B **12**, 3795 (1972).

[5] C.K. Majumdar and D.K. Ghosh, J. Math. Phys. **10**, 1399 (1969); K. Okamoto and K. Nomura, Phys. Lett. **169**A, 433 (1992).

[6] I. Affleck and E. Lieb, Lett. Math. Phys. **12**, 57 (1986).

[7] E.H. Lieb, T. Schultz and D.J. Mattis, Ann. Phys. **16**, 407 (1961).

[8] M. Oshikawa, M. Yamanaka and I. Affleck, Phys. Rev. Lett. **78**, 1984 (1997).

[9] S.R. White, Phys. Rev. B **53**, 52 (1996); S. Brehmer, H-J. Mikeska and U. Neugebauer, J. Phys. Cond. Matt. **8**, 7161 (1996); A.K. Kolezhuk and H-J. Mikeska, cond-mat/9701089 (1997).

[10] P.D. Sacramento and V.R. Vieira, Z. Phys. B **101**, 441 (1996).

[11] P.D. Sacramento and V.R. Vieira, J. Phys. Cond. Matt. **7**, 8619 (1995).

[12] A. Kolezhuk, R. Roth and U. Schollwock, Phys. Rev. Lett. **77**, 5142 (1996).

[13] F.D.M. Haldane, Phys. Rev. Lett. **60**, 635 (1988).

[14] B.S. Shastry, Phys. Rev. Lett. **60**, 639 (1988).

[15] F. Gebhard and D. Vollhardt, Phys. Rev. Lett. **59**, 1472 (1987).

[16] *Correlation Effects in Low-Dimensional Electron Systems*, ed. A. Okiji and N.

Kawakami, Springer-Verlag 1994.

[17] F.D.M. Haldane, Phys. Rev. Lett. **66**, 1529 (1991).

[18] F.D.M. Haldane, in ref. 13, p. 3.

[19] We thank Duncan Haldane for a helpful discussion and for suggesting this model as a likely candidate for gapless behavior.

[20] R. Egger and H. Schoeller, Phys. Rev. B **54**, 16337 (1996).

[21] K. Hallberg and R. Egger, cond-mat/9702162 (1997).

[22] E.R. Gagliano, E. Dagotto, A. Moreo and F.C. Alcaraz, Phys. Rev. B **34**, 1677 (1986).

[23] M.N. Barber, *Phase Transitions*, vol. 8, 146 ed. C. Domb and J. Lebowitz, New York, Academic 1983.

[24] H. Betsuyaku, Phys. Rev. B **34**, 8125 (1986).

[25] See for example, M. Henkel and G. Schutz, J. Phys. A **21**, 2617 (1988) (we thank the referee for calling our attention to this reference).

[26] For $\alpha < \alpha_s$ the energy per spin increases with the system size and for $\alpha > \alpha_s$ it decreases, where $\alpha_s \sim 1.693$.

TABLE I- Groundstate energy per spin and gap as a function of $N$ for $\alpha = 1.01, 2, 3$.

| $N$ | $\alpha = 1.01$ $-E_N/N$ | $\alpha = 1.01$ $Gap$ | $\alpha = 2$ $-E_N/N$ | $\alpha = 2$ $Gap$ | $\alpha = 3$ $-E_N/N$ | $\alpha = 3$ $Gap$ |
|---|---|---|---|---|---|---|
| 4  | 2.0595832 | 1.1118877 | 2.1589760 | 1.2337006 | 2.2976817 | 1.3702968 |
| 6  | 2.3595499 | 0.8899970 | 2.0523213 | 0.8480498 | 1.9179450 | 0.8541843 |
| 8  | 2.5811593 | 0.7363615 | 2.0005515 | 0.6540099 | 1.7672689 | 0.6402738 |
| 10 | 2.7580452 | 0.6359315 | 1.9707343 | 0.5353366 | 1.6920636 | 0.5232326 |
| 12 | 2.9056771 | 0.5616502 | 1.9516741 | 0.4545769 | 1.6490979 | 0.4491504 |
| 14 | 3.0325603 | 0.5042410 | 1.9385998 | 0.3957728 | 1.6222435 | 0.3979216 |
| 16 | 3.1439108 | 0.4584054 | 1.9291615 | 0.3509020 | 1.6043433 | 0.3603571 |

TABLE II- Extrapolated values of the gap using the VBS method.

| | $\alpha_m = -[1-(-1)^m]/2$ | $\alpha_m = 1$ | $\alpha_m = 0$ |
|---|---|---|---|
| $\alpha = 1.01$ | 0.0658 | 0.2004 | 0.1254 |
| $\alpha = 2.0$ | -0.0142 | 0.1491 | 0.0742 |
| $\alpha = 2.3$ | 0.0041 | 0.1571 | 0.0829 |
| $\alpha = 2.5$ | 0.0065 | 0.1677 | 0.0925 |
| $\alpha = 2.7$ | 0.0862 | 0.1850 | 0.1268 |
| $\alpha = 3.0$ | 0.1701 | 0.2202 | 0.1845 |

TABLE III- Extrapolation iterations for the gap for $\alpha = 2$ using the Hamer-Barber algorithm.

| $m$ | 0 | 1 | 2 | 3 |
|---|---|---|---|---|
| | 1.2337006 | | | |
| | 0.8480498 | 0.4575103 | | |
| | 0.6540099 | 0.3484716 | 0.1817683 | |
| | 0.5353366 | 0.2825508 | 0.1477233 | -0.0141768 |
| | 0.4545769 | 0.2382768 | 0.1244493 | |
| | 0.3957728 | 0.2064011 | | |
| | 0.3509020 | | | |

TABLE IV- Extrapolation iterations for the gap for $\alpha = 2.5$ using the Hamer-Barber algorithm.

| m | 0 | 1 | 2 | 3 |
|---|---|---|---|---|
|  | 1.3002061 |  |  |  |
|  | 0.8478680 | 0.4707043 |  |  |
|  | 0.6421958 | 0.3607076 | 0.1988730 |  |
|  | 0.5233554 | 0.2952211 | 0.1640630 | 0.0064759 |
|  | 0.4452184 | 0.2515429 | 0.1406424 |  |
|  | 0.3895432 | 0.2202066 |  |  |
|  | 0.3476438 |  |  |  |

TABLE V- Extrapolation iterations for the gap for $\alpha = 3$ using the Hamer-Barber algorithm.

| $m$ | 0 | 1 | 2 | 3 |
|---|---|---|---|---|
| | 1.3702968 | | | |
| | 0.8541843 | 0.4888593 | | |
| | 0.6402738 | 0.3818190 | 0.243082 | |
| | 0.5232326 | 0.3213966 | 0.216709 | 0.1700793 |
| | 0.4491504 | 0.2830859 | 0.202198 | |
| | 0.3979216 | 0.2570883 | | |
| | 0.3603571 | | | |

TABLE VI- Values of the error $\epsilon$ defined in eq. (7), for the several values of $\alpha$, calculated at the values of $\omega$ where the gap goes through zero (smaller than $1.0E-06$) and where $\epsilon$ is also small.

| $\alpha$ | $\omega$ | error ($\epsilon$) |
| --- | --- | --- |
| 1.01 | 0.846448 | 0.00068 |
| 2.0 | 0.962115 | 0.00017 |
| 2.3 | 0.890129 | 0.00066 |
| 2.5 | 0.768180 | 0.00158 |
| 2.7 | 0.590684 | 0.00005 |
| 3.0 | 0.534526 | 0.01318 |

TABLE VII- Minima of the error $\epsilon$ eq. (7) for the several values of $\alpha$ and the corresponding values of the gap. The minima of $\epsilon$ (smaller than $1.0E-05$) that are close to the set of values of $\omega$ shown in Table VI are highlighted in bold. The absolute value of these results for the gap are the error estimates within the method.

|  | $\omega_1$ | gap | $\omega_2$ | gap | $\omega_3$ | gap |
|---|---|---|---|---|---|---|
| $\alpha = 1.01$ | 0.810 | -0.0074257 | **0.864** | **0.0079885** | 1.077 | 0.0254297 |
| $\alpha = 2$ | 0.886 | -0.0067272 | **0.955** | **-0.0007323** | | |
| $\alpha = 2.3$ | 0.733 | -0.0239250 | **0.928** | **0.0034268** | | |
| $\alpha = 2.5$ | 0.622 | -0.0291923 | **0.883** | **0.0109872** | | |
| $\alpha = 2.7$ | 0.500 | -0.0163724 | **0.592** | **0.0002422** | 1.213 | 0.0713413 |
| $\alpha = 3$ | **0.914** | **0.1033700** | 1.445 | 0.1458258 | | |

TABLE VIII- Correlation functions $C(m)$ for $\alpha = 2$ for the set of values $N = 4$ to 16 with $N$ even.

| $N$ | 4 | 6 | 8 | 10 | 12 | 14 | 16 |
|---|---|---|---|---|---|---|---|
| $m$ | | | | | | | |
| 1 | -0.75000 | -0.70502 | -0.68749 | -0.67851 | -0.67318 | -0.66970 | -0.66728 |
| 2 | 0.50000 | 0.48340 | 0.47798 | 0.47560 | 0.47736 | 0.47363 | 0.47317 |
| 3 | | -0.55677 | -0.51869 | -0.50316 | -0.49500 | -0.49008 | -0.48683 |
| 4 | | | 0.45641 | 0.44601 | 0.44094 | 0.43810 | 0.43634 |
| 5 | | | | -0.47988 | -0.46342 | -0.45458 | -0.44917 |
| 6 | | | | | 0.43260 | 0.42565 | 0.42156 |
| 7 | | | | | | -0.44604 | -0.43647 |
| 8 | | | | | | | 0.41737 |

Figure Captions

Fig. 1- Gap as a function of $1/N$ for the standard frustrating interaction (positive) and Model (1) (alternated) for the values of $\alpha = 1.01, 2, 3$.

Fig. 2- Gap and estimate of the error as a function of $\omega$ for Model (1) for the values of $\alpha = 1.01, 2, 2.3, 2.5, 2.7, 3$.

Fig. 3- Gap and estimate of the error as a function of $\omega$ for the standard frustrating interaction (Haldane-Shastry model).

Fig. 4- a) Correlation functions $C(m)$ as a function of $m$ for $N = 16$ and $\alpha = 2$ for the Haldane-Shastry model (HS) and for Model (1). b) Correlation functions $C(m)$ for $m = 1, 2, 3, 4$ for $\alpha = 2$ as a function of $N$.

Fig. 5- Susceptibility as a function of temperature for $\alpha = 2$ for the values of $N = 4, 5, 6, 7, 8$.

Fig. 6- Correlation functions $C(m)$ as a function of temperature for the Haldane-Shastry model and for Model (1) for $N = 8$.

Fig. 7- Correlation function $C(N/2)$ as a function of temperature for $N = 8$ and $\alpha = 1.01, 2, 3$.